\documentclass[10pt]{amsart}
\usepackage{graphicx,color}
\usepackage{url}

\usepackage{mathrsfs}

\newtheorem{theorem}{Theorem}[section]
\newtheorem{lemma}[theorem]{Lemma}           
\newtheorem{cor}[theorem]{Corollary}
\newtheorem{prop}[theorem]{Proposition}

\theoremstyle{definition}
\newtheorem{definition}[theorem]{Definition}

\theoremstyle{remark}

\numberwithin{equation}{section}


\title[Generalized eigenfunctions for multi-dim. QWs]
{Asymptotic properties of generalized eigenfunctions for multi-dimensional quantum walks}
\author[T. Komatsu]{Takashi Komatsu}
\address[T. Komatsu]{Department of Bioengineering, School of Engineering,The University of Tokyo, Bunkyo-ku, Tokyo, 113-8656, Japan}
\email{komatsu@coi.t.u-tokyo.ac.jp}
\author[N. Konno]{Norio Konno}
\address[N. Konno]{Department of Applied Mathematics, Yokohama National University, Hodogaya, Yokohama, Kanagawa, 240-8501, Japan}
\email{konno-norio-bt@ynu.ac.jp}
\author[H. Morioka]{Hisashi Morioka}
\address[H. Morioka]{Graduate School of Science and Engineering,
Ehime University, Bunkyo-cho 3, Matsuyama, Ehime, 790-8577, Japan}
\email{morioka@cs.ehime-u.ac.jp}
\author[E. Segawa]{Etsuo Segawa}
\address[E. Segawa]{Graduate School of Environment and Information Sciences, Yokohama National University, Hodogaya, Yokohama, Kanagawa, 240-8501, Japan}
\email{segawa-etsuo-tb@ynu.ac.jp}
\thanks{H. Morioka is supported by the JSPS Grant-in-aid for young scientists No. 20K14327. 
E. Segawa is supported by the JSPS Grant-in-Aid for Scientific Research (C) No. 19K03616 and Research Origin for Dressed Photon.}

\date{\today}

\begin{document}
\maketitle

\begin{abstract} 

We construct a distorted Fourier transformation associated with the multi-dimensional quantum walk.
In order to avoid the complication of notations, almost all of our arguments are restricted to two dimensional quantum walks (2DQWs) without loss of generality.
The distorted Fourier transformation characterizes generalized eigenfunctions of the time evolution operator of the QW.
The 2DQW which will be considered in this paper has an anisotropy due to the definition of the shift operator for the free QW.
Then we define an anisotropic Banach space as a modified Agmon-H\"{o}rmander's $\mathcal{B}^*$ space and we derive the asymptotic behavior at infinity of generalized eigenfunctions in these spaces.
The scattering matrix appears in the asymptotic expansion of generalized eigenfunctions.
\end{abstract}

%
%
\section{Introduction}
\subsection{Scattering theory for multi-dimensional quantum walk}
In this paper, we consider the time-independent scattering theory for the position-dependent $d$-dimensional quantum walk ($d$DQW or simply QW for short) as a finite rank perturbation of the $d$D free QW which is defined as follows.
The states of quantum walker on ${\bf Z}^d = \{ x=(x_1,\ldots ,x_d ) \ ; \ x_1 , \ldots ,x_d \in {\bf Z} \} $ are represented by $ \psi = \{ \psi (x) \} _{x\in {\bf Z}^d} $ which are ${\bf C}^{2d} $-valued sequences.
Letting $*^{\mathsf{T}}$ be the transpose operator for matrices or vectors, we denote by the column vector
$$
\psi (x)=[ \psi^+ _{1} (x), \psi ^-_{1} (x), \ldots , \psi ^+_{d} (x), \psi ^-_{d} (x)]^{\mathsf{T}} ,\quad x\in {\bf Z}^d ,
$$
for ${\bf C}$-valued sequences $\psi ^{\pm}_{j}$ for $j=1,\ldots ,d$ on ${\bf Z}^d $.
Each component $\psi _{j}^{\pm} (x)$ corresponds the ``probability amplitude" at $x\in {\bf Z}^d$ of the chirality $(j,\pm)$. 
The \textit{$d$D free QW} is defined by the operator $U_0 =S$ where $S$ is the shift operator :
$$
(S\psi )(x) = [ \psi _{1}^+ (x+e_1), \psi _{1}^- (x-e_1 ), \ldots , \psi _{d}^+ (x+e_d) , \psi _{d}^- (x-e_d ) ] ^{\mathsf{T}} , 
$$
where $e_1,\ldots ,e_d$ are the standard basis on ${\bf Z}^d$.
The \textit{position-dependent QW} is defined by the operator $U=SC$ where $C$ is the operator of multiplication by a matrix $C(x)=( c_{k,l} (x) )_{k,l=1}^{2d} \in \mathrm{U} (2d)$ for every $x\in {\bf Z}^d $.
Throughout of this paper, we assume that the following statements hold.
\begin{itemize}

\item There exists a positive integer $ n_0$ such that $C(x)$ is the $2d \times 2d$ identity matrix $I_{2d}$ for $x\in {\bf Z}^d \setminus D$ where $ D=\{ x\in {\bf Z}^d \ ; \ |x_1 |\leq n_0 ,\ldots , \ |x_d|\leq n_0 \} $.

\item For every $x\in D$, 
$$
\mathrm{det} (c_{2k-1,2l-1} (x))_{k,l=1}^d , \quad \mathrm{det} (c_{2k,2l} (x))_{k,l=1}^d ,
$$
do not vanish.
\end{itemize}
These assumptions will be used in order to prove a unique continuation property for generalized eigenfunctions of $U$ in Lemma \ref{S3_lem_UCP} and Corollary \ref{S3_cor_UCP2}.
The unique continuation property guarantees a radiation condition for non-homogeneous equations and the absence of eigenvalues of $U$.
See Corollaries \ref{S3_cor_helmhltzunique} and \ref{S3_cor_absenceEV}.

In the following, almost all of our arguments are restricted to the case $d=2$ in order to avoid the complication of notations.
Our results can be generalized easily for higher dimensional cases. 
For the case $d=2$, we use more explicit notations.
Components of ${\bf C}^4$-valued sequences $\psi$ are written as 
$$
\psi (x)=[ \psi _{\leftarrow} (x), \psi _{\rightarrow} (x), \psi _{\downarrow} (x), \psi _{\uparrow} (x)]^{\mathsf{T}}, \quad x\in {\bf Z}^2 .
$$
The chirality of $\psi$ is represented by $p\in \{ \leftarrow , \rightarrow , \downarrow , \uparrow \}$.
Typical examples of $C(x)$ for every $x\in D$ such that the second assumption holds are
$$
\frac{1}{2} \left[ \begin{array}{cccc}
1 & 1 & 1 & 1 \\ 1 & -1 & 1 & -1 \\ 1 & 1 & -1 & -1 \\ 1 & -1 & -1 & 1 \end{array} \right] , \quad \left[ \begin{array}{cccc}
1/\sqrt{2} & 1/\sqrt{6} & 1/\sqrt{12} & 1/2 \\ -1 /\sqrt{2} & 1/\sqrt{6} & 1/\sqrt{12} & 1/2 \\ 0 & -2/\sqrt{6} & 1/\sqrt{12} & 1/2 \\ 0 & 0 & 3/\sqrt{12} & -1/2 \end{array} \right] . 
$$
On the other hand, the 2D Grover coin and the 2D Fourier coin
$$
 \frac{1}{2} \left[ \begin{array}{cccc}
-1 & 1 & 1 & 1 \\ 1 & -1 & 1 & 1 \\ 1 & 1 & -1 & 1 \\ 1 & 1 & 1 & -1 \end{array} \right] , \quad 
\frac{1}{2} \left[ \begin{array}{cccc}
1 & 1 & 1 & 1 \\ 1 & i & -1 & -i \\ 1 & -1 & 1 & -1 \\ 1 & -i & -1 & i \end{array} \right] ,
$$
do not satisfy the second assumption.

Obviously, operators $ U_0 $ and $ U$ are unitary on the Hilbert space $\ell^2 := \ell^2 ({\bf Z}^2 ; {\bf C}^4 )$ equipped with the standard inner product 
$$
(f,g)_{\ell^2} = \sum_{p\in \{ \leftarrow, \rightarrow , \downarrow , \uparrow \}} \sum _{x\in {\bf Z}^2} f_{p} (x)\overline{g_{p} (x)} .
$$
The discrete time evolutions of these QWs are given by 
$$
\Psi (t,\cdot )= U^t \psi , \quad \Psi ^{(0)} (t,\cdot )= U_0^t \psi , \quad t\in {\bf Z} ,
$$
for an initial state $\psi $.
If $\psi \in \ell^2 $, these time evolutions preserve the $\ell^2$-norm of $\psi$ i.e. $\| \Psi (t,\cdot )\| _{\ell^2} =\| \Psi^{(0)} (t,\cdot) \| _{\ell^2} =\| \psi \| _{\ell^2} $ for any $t\in {\bf Z} $.

\begin{figure}[t]
\centering
\includegraphics[bb=0 0 541 201, width=6cm]{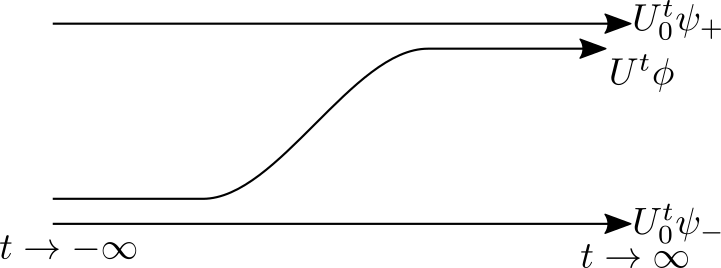}
\caption{For an initial state $\phi \in \mathcal{H}_{ac} (U)$, there exist $\psi_{\pm}\in \ell^2$ such that $U^t \phi \sim U_0^t \psi_{\pm}$ as $t\to \pm \infty$. The wave operators represent these asymptotic behaviors. Roughly speaking, the scattering in view of the dynamics of $U^t \phi$ is a transition from a free QW $U_0^t \psi_- $ to another free QW $U_0^t \psi_+ $. The scattering operator $\Sigma$ relates $\psi_-$ and $\psi_+$.}
\label{fig_scattering}
\end{figure}

In view of the quantum scattering theory, the scattering matrix (S-matrix) is an important object. 
There  are several equivalent definitions of the S-matrix.
In the time-dependent scattering theory, we can prove that the wave operators 
$$
W_{\pm} = {\mathop{{\rm s\text{-}lim}}_{t\to\pm \infty}} \, U^{-t} U_0^t  \quad \text{in} \quad \ell^2  
$$
exist and asymptotically complete under our assumption by the similar way of Suzuki \cite{Su}.
Namely, we have the following fact on the long time behavior of QWs.
See also Figure \ref{fig_scattering}.

\begin{prop}
The ranges of the wave operators coincide with $\mathcal{H}_{ac} (U)$, the absolutely continuous subspace for $U$.
In particular, for any $\phi \in \mathcal{H}_{ac} (U)$, there exist $\psi _{\pm} \in \ell^2 $ such that $\| U^t \phi    -U^t_0 \psi_{\pm} \|_{\ell^2} \to 0$ as $t\to \pm \infty $.
The wave operators are unitary on $\ell^2  $ and we have $ W_{\pm}^* = W_{\pm}^{-1} $.
\label{S1_prop_waveop}
\end{prop}

Note that this proposition holds under the short-range condition.
Namely, the first assumption for $ C(x)$ can be replaced by $\| C(x)-I_4 \| \leq c (1+|x|)^{-1-\epsilon}$ for some constants $c,\epsilon >0$ for Proposition \ref{S1_prop_waveop}.

The scattering operator $\Sigma = W_+^* W_- : \psi_- \mapsto \psi_+ $ is another important object.
In view of Proposition \ref{S1_prop_waveop}, $\Sigma$ connects the behavior of the quantum walker as $t\to -\infty $ to $t\to \infty $ in terms of the 2D free QW i.e. $\Sigma$ represents a transition from the 2D free quantum walker $\psi_-$ to another free quantum walker $\psi_+$.
The S-matrix $\widehat{\Sigma} (\theta )$ for $\theta \in [0,2\pi )$ is given by the spectral decomposition of $\widehat{\Sigma}$ which is a spectral transform of $\Sigma$ which will be introduced in the formula (\ref{S4_eq_spectraltransS}) as 
$$
\widehat{\Sigma} = \int _0^{2\pi} \oplus \widehat{\Sigma} (\theta )d\theta .
$$
For details of the time-dependent scattering theory for QWs, see Suzuki \cite{Su} or Morioka \cite{Mo}.
In these previous works, the authors consider 1DQWs.
However, their arguments rely on abstract operator theory and spectral theory for unitary operators.
Then it can be applied for our case easily.

\begin{figure}[b]
\centering
\includegraphics[bb=0 0 977 429, width=9cm]{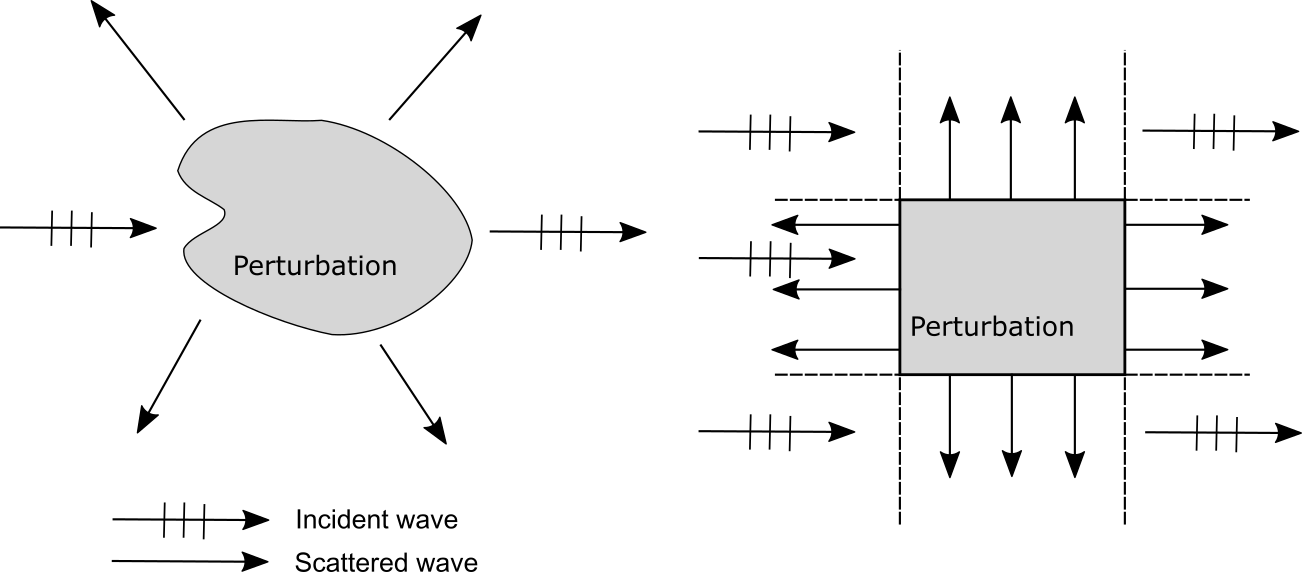}
\caption{The notion of the scattering in view of generalized eigenfunctions associated with the continuous spectrum. For Schr\"{o}dinger equations $ (-\Delta +V(x))u=\lambda u$ or acoustic wave equations $-c(x)^2 \Delta u=\lambda u$ on ${\bf R}^d$, the scattered wave is a spherical wave up to lower order terms (Left figure). However, for QWs on ${\bf Z}^d $, spherical waves do not appear, and the scattered wave passes along
corridors (Right figure). Entries of the S-matrix appear in the scattered wave as its amplitude and phase shift.}
\label{fig_tiscattering}
\end{figure}
The wave operators and the scattering operator are considered in the Hilbert space $\ell^2$. 
From the physical point of view, it is reasonable to consider plane waves and corresponding scattered waves in certain classes larger than $\ell^2$. 
Moreover, the S-matrix naturally appears in a spatial asymptotic behavior at infinity of a generalized eigenfunction to the equation
$$
(U-e^{i\theta} )u=0 \quad \text{on} \quad {\bf Z}^2 , \quad \theta \in [0,2\pi ).
$$
In view of the separation of variables $\Psi (t,x)=e^{it\theta} u(x)$ where $u$ satisfies the above equation, we have a time evolution $\Psi (t,\cdot )=U^t u$ for any $t\in {\bf Z}$. 
Note that generalized eigenfunctions do not belong to $\ell^2$ when $e^{i\theta}$ is included in the continuous spectrum of $U$.
Generalized eigenfunctions associated with a continuous spectrum consist of an incident plane wave and the corresponding scattered wave.
The S-matrix appears in the scattered wave as its amplitude and phase shift.
See Figure \ref{fig_tiscattering}.
Thus the spectral theory for $ U$ is an important topic in the research area of the scattering theory for QWs.

The main purpose of this paper is to construct and to characterize generalized eigenfunctions in an anisotropic Banach space on ${\bf Z}^2$.
The spectral theory for the unitary operator $U$ allows us to see the behavior of the outgoing scattered wave.
In order to study the distorted Fourier transformation, we often use the Green function which is the kernel of the resolvent operator $R_0 (\kappa )=(U_0 -e^{i\kappa} )^{-1} $, $\kappa \in {\bf C}\setminus {\bf R}$.
If $\kappa \in {\bf R}$, the operator $R_0 (\kappa )$ does not make sense in $\ell^2 $.
However, we will show that the limit $ R_0 (\theta \pm i0 )$ for $\theta \in {\bf R}$ exists in the sense of Agmon-H\"{o}rmander's $\mathcal{B}$-$\mathcal{B}^*$ spaces (\cite{AgHo}).

\subsection{Agmon-H\"{o}rmander's spaces}
Agmon-H\"{o}rmander's $\mathcal{B}$-$\mathcal{B}^*$ spaces are often used in the time-independent scattering theory for Schr\"{o}dinger equations.
Consider the equation 
$$
(-\Delta +V-\lambda )u=0 \quad \text{on} \quad {\bf R}^d , \quad \lambda >0,
$$
where $V\in C_0^{\infty} ({\bf R}^d)$ is a real-valued function.
The solution to this equation can be characterized by the Banach space $\mathcal{B}^*$ equipped with the norm
$$
\| u\|^2 _{\mathcal{B}^*} = \sup _{\rho >1} \frac{1}{\rho} \int _{|x|<\rho} |u(x)|^2 dx <\infty .
$$
The solution $u$ has the asymptotics 
$$
u(x)=|x|^{-(d-1)/2} \left(C(\lambda)e^{i\sqrt{\lambda}|x|} \phi_+ (\omega)  +\overline{C(\lambda)} e^{-i\sqrt{\lambda}|x|} \phi_- (\omega ) \right) +o(|x|^{-(d-1)/2 } )
$$
as $|x|\to \infty$ where $\omega =x/|x|$.
It is well-known that the S-matrix for Schr\"{o}dinger operators relates $\phi_+$ and $\phi_- $.
For this topic, see Yafaev \cite{Ya}.

For our 2DQW, it is inadequate to adopt the usual separation of variables and to derive the asymptotic expansion with respect to the radius due to anisotropy of $U$ and $U_0$.
Then we introduce a pair of anisotropic $\mathcal{B}$-$\mathcal{B}^*$ spaces.
We will derive the asymptotic behaviors of generalized eigenfunctions by evaluating functions for each direction depending on its chirality.

\subsection{Related works}

The scattering theory for one dimensional QWs has been studied in some previous works.
Feldman-Hillery \cite{FeHi1,FeHi2} are pioneering studies of QWs in view of the scattering theory. 
As has been mentioned above, Suzuki \cite{Su} proved the existence and the asymptotic completeness of the wave operators.
Richard et al. \cite{RST,RST2} considered more general cases.
Note that the authors adopted commutator method for unitary operators (see \cite{ABMG}, \cite{FRT}, \cite{Sh}).
Morioka \cite{Mo}, Morioka-Segawa \cite{MoSe}, Maeda et al. \cite{MS42} and Komatsu et al. \cite{KKMS} considered the time-independent scattering theory and the absence of eigenvalues embedded in the continuous spectrum.
Tiedra de Aldecoa \cite{Ti} studied an abstract theory of time-independent scattering for unitary operators and its applications to QWs.
Maeda et al. \cite{MS41} solved an inverse scattering problem for a nonlinear QW.

Our method adopted in this paper is an analogue of the time-independent scattering theory for self-adjoint Hamiltonians.
Over the past few years, the time-independent scattering theory for discrete Schr\"{o}dinger operators has been studied.
The following works are deeply related with our arguments: Isozaki-Korotyaev \cite{IsKo}, Isozaki-Morioka \cite{IsMo,IsMo2}, Ando et al. \cite{AIM,AIM2}.
Nakamura \cite{Na} constructed the S-matrix as a pseudo-differential operator for discrete Schr\"{o}dinger operators (as one of examples of applications) by using the microlocal analysis.
Tadano \cite{Ta} studied the scattering theory for discrete Schr\"{o}dinger operators with long-range perturbations.

\subsection{Results and plan of this paper}

In Section 2, we introduce some functional spaces.
The anisotropic $\mathcal{B}$-$\mathcal{B}^*$ spaces are defined here.
Some properties of these functional spaces will be proven.

In Section 3, we study some properties of the Green function associated with the 2D free QW.
By using the Green function of the 1D free QW, we can derive an explicit formula of the Green function of the 2D free QW.
Then we can see the asymptotic behavior of the solution in $\mathcal{B}^*$ to the equation $(U_0 -e^{i\theta} )u=f$ for $f\in \mathcal{B}$.
In view of this asymptotics, we can define the radiation condition which guarantees the uniqueness of the solution to the equation $(U_0 -e^{i\theta} )u=f$.
Here we apply a multi-dimensional generalization of the unique continuation property for QWs (Corollary \ref{S3_cor_UCP2}).
As a consequence of the argument in Section 3, we have the absence of eigenvalues of $U$ (Corollary \ref{S3_cor_absenceEV}).

In Section 4, we consider the generalized eigenfunction for $U$.
We introduce a combinatorial construction of generalized eigenfunctions in $\mathcal{B}^* $.
This construction is a multi-dimensional version of the result in \cite{KKMS}. 
This approach is based on the long time limit of the dynamics of QWs.
After that, we discuss the spectral theory for $U$ in order to characterize rigorously the set of generalized eigenfunctions in $\mathcal{B}^* $.
The existence of the limits $R(\theta \pm i0 ) $ in $\mathcal{B}$-$\mathcal{B}^*$ spaces for $R(\kappa )=(U-e^{i\kappa} )^{-1}$ is proved here.
The spectral representation of $U$ is introduced as a distorted Fourier transformation associated with $U$.
Due to the closed range theorem, we prove a characterization of generalized eigenfunctions in $\mathcal{B}^*$ (Theorem \ref{S4_thm_chsolution}).
Finally, we show that the S-matrix $\widehat{\Sigma} (\theta )$ naturally appears in the asymptotic behavior of generalized eigenfunctions.
We observe that the scattered wave does not spread radially but passes along some corridors (Theorem \ref{S4_thm_corridor}).

\subsection{Notation}

The notation used throughout of this paper is as follows.
${\bf T}^2 := {\bf R}^2 /(2\pi {\bf Z}^2 ) $ denotes the flat torus.
For $ u,v\in {\bf C}^n $, $\langle u,v \rangle $ denotes the inner product of ${\bf C}^n$, and we put $|u|=\sqrt{ \langle u,u \rangle} $.
For a matrix $A$, $A^*$ denotes the Hermitian conjugate $\overline{A^{\mathsf{T}}}$.
$\mathrm{diag} [ a_1 , \ldots , a_n ]$ denotes the $n\times n$ diagonal matrix.

For a ${\bf C}^4 $-valued sequence $f=\{ f(x) \} _{x\in {\bf Z}^2} $, the mapping $\mathcal{U}$ is the Fourier transformation
$$
\widehat{f} (\xi ) := (\mathcal{U} f)(\xi )=\frac{1}{2\pi} \sum _{x\in {\bf Z}^2} e^{-i \langle x,\xi \rangle } f(x), \quad \xi \in {\bf T}^2 .
$$
The Fourier coefficients of a distribution $\widehat{g} $ on ${\bf T}^2 $ is given by 
$$
g(x):= (\mathcal{U}^* \widehat{g} )(x)= \frac{1}{2\pi} \int _{{\bf T}^2} e^{i \langle x,\xi \rangle} \widehat{g} (\xi ) d\xi , \quad x\in {\bf Z}^2 .
$$

For Banach spaces $X$ and $Y$, ${\bf B} (X;Y)$ denotes the set of bounded linear operators from $X$ to $Y$.
For an operator $L \in {\bf B} (\ell^2 ;\ell^2 )$, $L^*$ is the adjoint operator of $L$ with respect to the inner product of $\ell^2 $.
As usual, for an operator $L\in {\bf B} (X;Y) $ for Banach spaces $X$ and $Y$, $L^*$ is its adjoint operator in ${\bf B} (Y^* ;X^*)$.
However, for $U=SC$ and $U_0 =S$, we often adopt the abuse of notations $U^* = C^* S^{-1}$ and $U_0^* = S^{-1} $ on some Banach spaces.


\section{Functional spaces}
\subsection{Anisotropic Agmon-H\"{o}rmander spaces}

The Banach spaces $\mathcal{B}$ and $\mathcal{B}^* $ which are defined here are used for the proof of boundedness of $R_0 (\kappa )=(U_0 -e^{i\kappa} )^{-1} $ and $ R(\kappa )= (U- e^{i\kappa} )^{-1} $ with $\kappa \in {\bf C} \setminus {\bf R} $.
Let $r_{-1} =0$, $r_j =2^j $ for $j\geq 0$, and $ I_j = \{ y\in {\bf Z} \ ; \ r_{j-1} \leq |y| < r_j \} $.
For a ${\bf C}^4 $-valued sequence $f=\{ f(x) \} _{x \in {\bf Z}^2} $, we put 
$$
a_j (f) ^2 =  \sum _{p\in \{ \leftarrow , \rightarrow \}}  \sum_{x_1 \in I_j} \sum _{x_2 \in {\bf Z}}  |f_{p} (x)|^2     + \sum_{p\in \{ \downarrow, \uparrow \}}    \sum_{x_2 \in I_j} \sum _{x_1 \in {\bf Z}} |f_{p} (x)|^2  .
$$
The Banach space $ \mathcal{B} $ is defined by  
$$
\mathcal{B} = \left\{ f=\{ f(x)\} _{x\in {\bf Z}^2} \ ; \ \| f\| _{\mathcal{B}} = \sum _{j=0}^{\infty} r_j^{1/2} a_j (f) <\infty \right\} .
$$

\begin{lemma}
The dual space $ \mathcal{B}^* $ is equipped with the norm 
$$
\| u\| _{\mathcal{B}^*} = \sup _{j\geq 0} r_j^{-1/2} a_j (u).
$$
\label{S2_lem_bdual}
\end{lemma}

Proof.
We take $f^{(k)} \in \mathcal{B}$ such that $\mathrm{supp} f_{\leftarrow}^{(k)} , \, \mathrm{supp} f_{\rightarrow}^{(k)} \subset \{ x\in {\bf Z}^2 \ ; \ r_{k-1} \leq |x_1 | <r_k \} $ and $\mathrm{supp} f_{\downarrow}^{(k)} , \, \mathrm{supp} f_{\uparrow}^{(k)} \subset \{ x\in {\bf Z}^2 \ ; \ r_{k-1} \leq |x_2 | <r_k \} $.
For $T\in \mathcal{B}^* $, we consider the restriction $T^{(k)}  $ on the subspace $\mathcal{H}_k := \mathcal{V}_k \oplus \mathcal{W}_k$ where
\begin{gather*}
\mathcal{V}_k = \{ f\in \ell^2 (I_k \times {\bf Z} ; {\bf C}^4 ) \ ; \ f(x)=[ f_{\leftarrow} (x) , f_{\rightarrow} (x), 0 ,0]^{\mathsf{T}} \} , \\
\mathcal{W}_k = \{ f\in \ell^2 ({\bf Z} \times I_k  ; {\bf C}^4 ) \ ; \ f(x)=[ 0,0,f_{\downarrow} (x) , f_{\uparrow} (x)]^{\mathsf{T}} \} .
\end{gather*}
Note that $\mathcal{H}_k$ for every $k$ is a Hilbert space.
We have 
$$
| T^{(k)} (f^{(k)} )| \leq \| T \| \| f^{(k)} \| _{\mathcal{B}} = \| T\| r_k^{1/2} a_k (f^{(k)} ) =\| T\| \| f^{(k)} \| _{\mathcal{H}_k} r_k^{1/2}.
$$
Applying the Riesz representation theorem on $\mathcal{H}_k$, we see that there exists $u^{(k)} \in \mathcal{H}_k $ such that 
\begin{gather}
\begin{split}
&T^{(k)} (f^{(k)} ) \\
& = \sum_{p\in \{ \leftarrow , \rightarrow \}}  \sum _{x_1 \in I_k} \sum _{x_2 \in {\bf Z}}   f^{(k)} _{p} (x) \overline{u^{(k)} _{p} (x)}   +  \sum_{p\in \{ \downarrow , \uparrow \}} \sum _{x_2 \in I_k} \sum _{x_1 \in {\bf Z}}   f^{(k)} _{p} (x) \overline{u^{(k)} _{p} (x)} , 
\end{split}
\label{S2_eq_tkfk}
\end{gather}
and
$$
\| u^{(k)} \|_{\mathcal{H}_k} = a_k (u^{(k)} )= \sup _{\| f^{(k)} \| _{\mathcal{H}_k} =1 } | T^{(k)} (f^{(k)} )|  \leq \| T\| r_k^{1/2} .
$$
Taking a sequence $u$ such that $u=u^{(k)} $ on every subspace $\mathcal{H}_k$, we have 
$$
\sup _{j\geq 0} r_j ^{-1/2 } a_j (u) \leq \| T\| ,
$$ 
which implies $\| u\| _{\mathcal{B}^*} \leq \| T\|$.

For any $ f\in \mathcal{B}$, we can write $f=\sum _{k=0}^{\infty} f^{(k)}$ for $f^{(k)}\in \mathcal{H}_k$.
In fact, we see 
$$
f(x)= \sum_{k=0}^{\infty} \left( \chi _{\mathcal{V}_k} (x) \left[ \begin{array}{c} f_{\leftarrow}(x) \\ f_{\rightarrow} (x) \\ 0 \\ 0 \end{array} \right] + \chi _{\mathcal{W}_k} (x) \left[ \begin{array}{c} 0 \\ 0 \\  f_{\downarrow}(x) \\ f_{\uparrow} (x) \end{array} \right] \right) , \quad x\in {\bf Z}^2 ,
$$
where $\chi _{\mathcal{V}_k} $ and $\chi _{\mathcal{W}_k} $ are characteristic functions of $I_k \times {\bf Z}$ and ${\bf Z} \times I_k$, respectively. 
In view of the definition of $\| \cdot \|_{\mathcal{B}} $, we have  
$$
f^{(k)}: =\chi _{\mathcal{V}_k} \left[ \begin{array}{c} f_{\leftarrow} \\ f_{\rightarrow}  \\ 0 \\ 0 \end{array} \right] + \chi _{\mathcal{W}_k} \left[ \begin{array}{c} 0 \\ 0 \\  f_{\downarrow} \\ f_{\uparrow}  \end{array} \right] \in \mathcal{H}_k .
$$
 We have
$$
T(f)= \sum_{k=0}^{\infty} T^{(k) } (f^{(k)} ).
$$
It follows from (\ref{S2_eq_tkfk}) that 
$$
|T(f)| \leq \sum_{k=0}^{\infty} r_k^{1/2} a_k (f) r_k^{-1/2} a_k (u) \leq \left( \sup_{j\geq 0} r_j^{-1/2} a_j (u) \right) \sum _{k=0}^{\infty} r_k^{1/2} a_k (f).
$$
This inequality implies 
$$
\| T\| \leq \sup _{j\geq 0} r_j^{-1/2} a_j (u) =\| u\| _{\mathcal{B}^*} .
$$
Therefore, we obtain $\| T\| = \| u\| _{\mathcal{B}^*} $.
Since $T\in \mathcal{B}^*$ is arbitrary, we have proven the lemma.
\qed

\medskip

The following equivalent norm of $ \mathcal{B}^* $ is easier to handle :
\begin{gather}
\begin{split}
&M _{\mathcal{B}^*} (u)^2 \\
&= \sup_{\rho >1} \frac{1}{\rho} \left( \sum_{p\in \{ \leftarrow , \rightarrow \}} \sum_{|x_1|<\rho} \sum_{x_2 \in {\bf Z}} |u_{p} (x)|^2  +\sum_{p\in \{ \downarrow , \uparrow \}} \sum_{|x_2|<\rho} \sum_{x_1 \in {\bf Z}}  |u_{p} (x)|^2 \right) .
\end{split}
\label{S2_eq_normB*}
\end{gather} 

\begin{lemma}
There exist constants $c_2 \geq c_1 >0$ such that
$$
c_1 \| u\|^2 _{\mathcal{B}^*} \leq M_{\mathcal{B}^*} (u)^2 \leq c_2 \| u\|^2 _{\mathcal{B}^*} .
$$
Then $\| \cdot \| _{\mathcal{B}^*} $ and $M_{\mathcal{B}^*} (\cdot )$ are equivalent as the norms on $\mathcal{B}^*$.
\label{S2_lem_equivalentnorm}
\end{lemma}

Proof.
For any $\epsilon >0$, there exists a nonnegative integer $k$ such that 
$$
\| u\|^2 _{\mathcal{B}^*} \leq r_k^{-1} a_k (u)^2 +\epsilon .
$$
Letting $ \rho = r_k$, we have 
\begin{gather*}
\begin{split}
&\| u\|^2 _{\mathcal{B}^*} \\
&\leq \frac{1}{\rho} \left( \sum_{p\in \{ \leftarrow , \rightarrow \}} \sum_{\rho /2 \leq |x_1|<\rho} \sum_{x_2 \in {\bf Z}} |u_{p} (x)|^2 +  \sum_{p\in \{ \downarrow , \uparrow \}} \sum_{\rho /2 \leq |x_2|<\rho} \sum_{x_1 \in {\bf Z}} |u_{p} (x)|^2 \right) +\epsilon \\
&\leq M_{\mathcal{B}^*} (u) ^2 +\epsilon .
\end{split}
\end{gather*}

On the other hand, for any $ \epsilon >0 $, there exists $\rho_0 >1$ such that 
\begin{gather*}
\begin{split}
&M_{\mathcal{B}^*} (u)^2 \\
&\leq \frac{1}{\rho_0} \left( \sum_{p\in \{ \leftarrow , \rightarrow \}} \sum_{ |x_1|<\rho_0} \sum_{x_2 \in {\bf Z}} |u_{p} (x)|^2  + \sum_{p\in \{ \downarrow , \uparrow \}} \sum_{ |x_2|<\rho_0} \sum_{x_1 \in {\bf Z}} |u_{p} (x)|^2 \right) +\epsilon .
\end{split}
\end{gather*}
Taking a positive integer $k$ such that $r_{k-1} \leq \rho_0 \leq r_k $, we have 
$$
M_{\mathcal{B}^*} (u)^2 \leq r_{k-1}^{-1} \sum_{j=0}^k a_j (u)^2 +\epsilon \leq \left(  \sup_{j\geq 0} r_j^{-1/2} a_j (u) \right)^2 r_{k-1}^{-1} \sum_{j=0}^k r_j +\epsilon .
$$
Thus we obtain
$$
M_{\mathcal{B}^*} (u)^2  \leq c\| u\|^2_{\mathcal{B}^*} +\epsilon ,
$$
for a constant $c>0$.
\qed

\medskip

In the following, we often use the norm $\| u\|_{\mathcal{B}^*} = M_{\mathcal{B}^*} (u)$ i.e. 
$$
\mathcal{B}^* = \left\{ u=\{ u(x)\} _{x\in {\bf Z}^2} \ ; \ M_{\mathcal{B}^*} (u) <\infty \right\} .
$$

Let the Hilbert space $\ell^{2,s} $ for $s\in {\bf R}$ be defined by the norm
$$
\| f\| ^2 _{\ell^{2,s}} = \sum_{ p\in \{ \leftarrow , \rightarrow \}} \sum _{x\in {\bf Z}^2} (1+|x_1|^2 )^s |f_{p} (x)|^2  +\sum_{ p\in \{ \downarrow , \uparrow \}} \sum _{x\in {\bf Z}^2} (1+|x_2|^2 )^s  |f_{p} (x)|^2  .
$$
When $s=0$, $\ell^{2,0 } =\ell^2 $ is the usual $\ell^2$-space equipped with the standard inner product.

In the following, $(u,f)$ denotes the pairing
$$
(u,f)= \sum _{x\in {\bf Z}^2} \langle u(x),f(x)\rangle ,
$$
for $u\in \mathcal{B}^*$ and $f\in \mathcal{B}$ or $u\in \ell^{2,-s} $ and $f\in \ell^{2,s}$ for $s\geq 0$.

Finally, we define the subspace $\mathcal{B}_0^*$ as the totality of sequences $u\in \mathcal{B}^*$ such that 
$$
\lim _{\rho \to 0} \frac{1}{\rho}  \left( \sum_{ p\in \{ \leftarrow , \rightarrow \}} \sum_{|x_1|<\rho} \sum_{x_2 \in {\bf Z}} |u_{p} (x)|^2 + \sum_{ p\in \{ \downarrow , \uparrow \}} \sum_{|x_2|<\rho} \sum_{x_1 \in {\bf Z}}  |u_{p} (x)|^2 \right) =0.
$$
When $u-v\in \mathcal{B}_0^*$, we write $u\simeq v$.
Then the subspace $\mathcal{B}_0^*$ is written as 
$$
\mathcal{B}_0^* = \{ u\in \mathcal{B}^* \ ; \ u\simeq 0 \} .
$$


\subsection{Some properties}
The following inclusion relation holds.

\begin{lemma}
We have
$$
\ell^{2,s} \subset \mathcal{B} \subset \ell^{2,1/2} \subset \ell^2 \subset \ell^{2,-1/2} \subset \mathcal{B}^* \subset \ell^{2,-s} ,
$$
for any $s>1/2$.
\label{S2_lem_inclusion}
\end{lemma}

Proof.
We put 
\begin{gather*}
\begin{split}
b_{j,s} (f)^2 = & \, \sum_{ p\in \{ \leftarrow , \rightarrow \}} \sum _{x_1 \in I_j} \sum _{x_2 \in {\bf Z}} (1+|x_1|^2 )^s |f_{p} (x)|^2  \\ 
& \, + \sum_{ p\in \{ \downarrow , \uparrow \}} \sum _{x_2 \in I_j} \sum _{x_1 \in {\bf Z}} (1+|x_2|^2 )^s |f_{p} (x)|^2 ,
\end{split}
\end{gather*}
for $j\geq 0$, $s\in {\bf R}$, and a sequence $f=\{ f(x)\} _{x\in {\bf Z}^2}$.
Note that $\| f\|^2 _{\ell^{2,s}} = \sum_{j=0}^{\infty} b_{j,s} (f)^2 $.
Since we have $b_{j,0} (f)= a_j (f)$, we also have $\| f\| _{\mathcal{B}} = \sum _{j=0}^{\infty} r_j^{1/2} b_{j,0} (f)$.
If $y\in I_j$, we can take a positive constant $c>0$ such that $c^{-1} r_j^2 \leq 1+|y|^2 \leq c r_j^2 $.
This inequality implies 
\begin{equation}
c^{-1} r_j^{2s} a_j (f)^2 \leq b_{j,s} (f) ^2 \leq c r_j^{2s} a_j (f)^2 ,
\label{S2_eq_ajbj}
\end{equation}
for every $j$ and $s \geq 0$.

For $s=1/2$, it follows from (\ref{S2_eq_ajbj}) that 
$$
\| f\| _{\ell^{2,1/2}} = \left( \sum _{j=0}^{\infty} b_{j,1/2} (f)^2 \right)^{1/2} \leq \sum _{j=0}^{\infty} b_{j,1/2} (f) \leq c\sum_{j=0}^{\infty} r_j^{1/2} a_j (f) \leq c\| f\| _{\mathcal{B}} ,
$$
for a constant $c>0$.
We obtain $\mathcal{B} \subset \ell^{2,1/2} $.

For $s>1/2 $, we use (\ref{S2_eq_ajbj}) again in order to show 
\begin{gather*}
\begin{split}
\| f\| _{\mathcal{B}} & = \sum_{j=0}^{\infty} r_j^{-s+1/2} r_j^s a_j (f) \leq \left( \sum_{j=0}^{\infty} r_j^{-2s+1} \right)^{1/2} \left( \sum_{j=0}^{\infty} r_j^{2s} a_j (f)^2 \right)^{1/2}  \\
& \leq c \| f\| _{\ell^{2,s} },
\end{split}
\end{gather*}
for a constant $c>0$.
Thus we have $ \ell^{2,s} \subset \mathcal{B}$.

Now we obtain $\ell^{2,s} \subset \mathcal{B} \subset \ell^{2,1/2} $ for $s>1/2$. 
Passing to the dual spaces, we also have $\ell^{2,-1/2} \subset \mathcal{B}^* \subset \ell^{2,-s}$.
\qed

\medskip

\textit{Remark.}
In view of Lemma \ref{S2_lem_inclusion}, the $\mathcal{B}$-$\mathcal{B}^*$ spaces constitute the optimal pair of Banach spaces to prove the limiting absorption principle for $R_0 (\kappa ) =(U_0 -e^{i\kappa} )^{-1}$ and $ R(\kappa )= (U-e^{i\kappa} )^{-1}$.
Namely, $\mathcal{B}$-$\mathcal{B}^*$ estimates are sharper than $\ell^{2,s}$-$\ell^{2,-s}$ estimates for $s>1/2$.
For details, we discuss in Sections 3 and 4.

\medskip

Let us show the following property.

\begin{lemma}
Suppose $f\in \mathcal{B}$.
For any $x_1, x_2 \in {\bf Z}$, we have $ f_{\leftarrow} (\cdot ,x_2 ) $, $f_{\rightarrow} (\cdot ,x_2 ) $, $ f_{\downarrow} (x_1 ,\cdot )$, $ f_{\uparrow} (x_1 , \cdot )\in \ell^1 ({\bf Z} )$.
\label{S2_lem_l1prop}
\end{lemma}

Proof.
The summability of $| f_{\leftarrow} ( y_1 , x_2 )|$ with respect to the variable $y_1$ follows from the estimate 
\begin{gather*}
\begin{split}
\sum _{y_1 \in {\bf Z}} | f_{\leftarrow} (y_1 , x_2 )| &\leq \sum _{j=0}^{\infty} r_j^{1/2} \left( \sum _{y_1 \in I_j} | f_{\leftarrow} (y_1 , x_2 )|^2   \right)^{1/2}  \\
& \leq \sum _{j=0}^{\infty} r_j^{1/2} \left( \sum _{y_1 \in I_j} \sum_{y_2 \in {\bf Z}} | f_{\leftarrow} (y_1 ,y_2 )|^2  \right)^{1/2} \leq \| f\| _{\mathcal{B}} .
\end{split}
\end{gather*}
The other cases can be proved in the same way.
\qed


\section{Radiation condition}

\subsection{Continuous spectrum}
The classification of the spectrum of a unitary operator is a consequence of the spectral theory for self-adjoint operators (see e.g. \cite{Ya2}).
In the following, $\sigma (A)$ is the totality of the spectrum of $A$.
There are two kinds of classification of $\sigma (A)$.
One is based on the spectral measure of $A$.
Namely, there exists a spectral decomposition $E_A (\theta )$ for $\theta \in {\bf R}$ such that $A$ can be represented by
$$
A= \int_0^{2\pi} e^{i\theta} dE_A (\theta ),
$$
where $E_A (\theta )=0$ for $\theta <0$ and $E_A (\theta )=1 $ for $\theta \geq 2\pi $.
Since $E_A (\theta )$ is a measure on ${\bf R} $, it provides a orthogonal decomposition of $\mathcal{H}$ as 
$$
\mathcal{H} = \mathcal{H}_p (A) \oplus \mathcal{H} _{ac} (A) \oplus \mathcal{H}_{sc} (A),
$$
where $\mathcal{H}_p (A)$ is the closure of the subspace spanned by eigenfunctions in $\mathcal{H}$ of $A$, $\mathcal{H}_{ac} (A)$ and $\mathcal{H}_{sc} (A )$ are orthogonal projections onto the absolutely continuous subspace and the singular continuous subspace with respect to the measure $E_A (\theta )$, respectively.
Then the spectrum $\sigma (A)$ is classified as 
\begin{gather*}
\sigma_p (A)= \{ \text{eigenvalues of } A \} ,\\
\sigma_{ac} (A)= \sigma (A| _{\mathcal{H}_{ac}} ), \quad \sigma _{sc} (A)= \sigma (A| _{\mathcal{H}_{sc}} ).
\end{gather*}
We call them the point spectrum, the absolutely continuous spectrum, and the singular continuous spectrum, respectively.

Another classification of $\sigma (A)$ is based on the topological point of view.
The discrete spectrum $\sigma_d (A)$ is the set of isolated eigenvalues of $A$ with finite multiplicities.
The essential spectrum $\sigma_{ess} (A)$ is defined by $ \sigma_{ess} (A)= \sigma (A) \setminus \sigma_d (A)$ i.e. $\sigma_{ess} (A)$ is the set of accumulation points in $\sigma (A)$.
Note that eigenvalues of infinite multiplicity are included in $\sigma_{ess} (A)$.

For the spectral theory for 2DQWs, we take $\mathcal{H}=\ell^2$ and $A=U_0$ or $U$.
First of all, let us derive the structure of $\sigma (U_0)$ explicitly.
In order to do this, we consider $\widehat{U}_0 = \mathcal{U} U_0 \mathcal{U}^*$ which is the operator of multiplication by the unitary matrix
$$
\widehat{U}_0 (\xi )= \mathrm{diag} [e^{i\xi_1}, e^{-i\xi_1} , e^{i\xi_2} , e^{-i\xi_2}]  , \quad \xi \in {\bf T}^2 .
$$
Obviously, we have 
$$
p(\xi ;\kappa ):= \det (\widehat{U}_0 (\xi )-e^{i\kappa})= \prod_{j=1}^2 (e^{i\xi_j} -e^{i\kappa} )(e^{-i\xi_j} -e^{i\kappa} ),
$$
for $\kappa \in {\bf C}$.
Then we obtain the spectrum $\sigma (U_0)$.

\begin{lemma}
We have $\sigma (U_0)= \sigma_{ess} (U_0)= \sigma_{ac} (U_0) =\{ e^{i\theta} \ ; \ \theta \in [0,2\pi )\} $.
\label{S3_lem_spec0}
\end{lemma}

Proof.
Due to the formula of $p(\xi ,\kappa )$, $\sigma_p (U_0)=\emptyset $ is trivial.
Let $ R_0 (\kappa )=(U_0 -e^{i\kappa} )^{-1} $ for $\kappa \in {\bf C} \setminus {\bf R}$.
In order to compute the spectral measure $E_{U_0} ((a,b)):=E_{U_0} (b-0)-E_{U_0} (a) $, we apply the formula 
\begin{equation}
(E_{U_0} ((a,b))f,f)_{\ell^2} = \lim_{\epsilon \downarrow 0} \int_a^b \frac{e^{i\theta}}{2\pi} (R_0 (\kappa_{+,\epsilon})f -R_0 (\kappa _{\epsilon,-} )f,f)d\theta , \quad f\in \ell^2, 
\label{S3_eq_stone}
\end{equation}
for $a<b$ and $\kappa _{\epsilon ,\pm } = \theta -i\log (1\mp \epsilon ) $.
For the proof of (\ref{S3_eq_stone}), see \cite[Lemma 4.5]{Mo}.
Since we have 
\begin{gather*}
\begin{split}
&\widehat{R}_0 (\kappa _{\epsilon ,\pm} ) \\
&= \mathrm{diag} [ (e^{i\xi_1} -e^{i\kappa_{\epsilon,\pm} })^{-1} , (e^{-i\xi_1} -e^{i\kappa_{\epsilon,\pm} })^{-1} ,(e^{i\xi_2} -e^{i\kappa_{\epsilon,\pm} })^{-1}  ,(e^{-i\xi_2} -e^{i\kappa_{\epsilon,\pm} })^{-1} ],
\end{split}
\end{gather*}
we can obtain 
\begin{gather*}
\begin{split}
(E_{U_0} ((a,b))f,f)_{\ell^2} &= \int _{a<\xi_1 <b} |\widehat{f}_{\leftarrow} (\xi )|^2 d\xi + \int _{a<-\xi_1 <b} |\widehat{f}_{\rightarrow} (\xi )|^2 d\xi \\ 
&\quad + \int _{a<\xi_2 <b} |\widehat{f}_{\downarrow} (\xi )|^2 d\xi + \int _{a<-\xi_2 <b} |\widehat{f}_{\uparrow} (\xi )|^2 d\xi .
\end{split}
\end{gather*}
Note that this formula follows from 
\begin{gather*}
\lim_{\epsilon \downarrow 0} \int_a^b \left( \frac{1}{e^{i(\omega -\theta )} -1+\epsilon} -\frac{1}{e^{i(\omega -\theta )} -1-\epsilon} \right) d\theta = \left\{ 
\begin{split}
2\pi ,& \quad \omega \in (a,b) , \\
0 , &  \quad \text{otherwise} ,
\end{split}
\right.
\end{gather*}
which can be proved by the complex contour integration in the similar way of \cite[Appendix A]{Mo} or \cite[Appendix A]{KKMS}.
Then the spectral measure $E_{U_0} (\theta )$ is absolutely continuous.
\qed

\medskip

As a consequence of Weyl's singular sequence method, we also determine $\sigma_{ess} (U)$ since $U-U_0$ is compact in $\ell^2$.

\begin{lemma}
We have $\sigma_{ess} (U)=\sigma_{ess} (U_0)=\{ e^{i\theta} \ ; \ \theta \in [0,2\pi )\} $.
As a consequence, we have $ \sigma_d (U)= \emptyset $.
\label{S3_lem_essspecu}
\end{lemma}

Proof.
A rigorous proof was given in \cite[Lemma 2.1]{MoSe} which is an analogue of well-known Weyl's singular sequence method on preservation of the essential spectrum for self-adjoint operators (See e.g. \cite[Section 0.3]{Ya2}).
\qed

\subsection{Green function and resolvent operator}

The resolvent operators 
$$
R_0 (\kappa )=(U_0 -e^{i\kappa} )^{-1} , \quad R(\kappa )=(U-e^{i\kappa} )^{-1},
$$
do not make sense as operators in ${\bf B} (\ell^2 ; \ell^2 )$ when $e^{i\kappa} \in \sigma (U_0)$ or $\sigma (U)$.
The limiting absorption principle ensure the existence of the limits $R_0 (\theta \pm i0)= \lim_{\epsilon \downarrow 0} R_0 (\theta-i\log (1\mp \epsilon ))$ and $R (\theta \pm i0)= \lim_{\epsilon \downarrow 0} R (\theta-i\log (1\mp \epsilon ))$ in ${\bf B} (\mathcal{B};\mathcal{B}^* )$ for $\theta \in [0,2\pi )$.
Here we consider $R_0 (\theta \pm i0 )$ by using an explicit formula of the Green function.

Now we seek a solution to the equation
\begin{equation}
(U_0 -e^{i\kappa} )u=f , \quad f\in \mathcal{B},
\label{S3_eq_helmholtz}
\end{equation}
for $\kappa \in {\bf C}\setminus {\bf R} $ such that $u$ is given by
$$
u(x)=\sum_{y\in {\bf Z}^2} G_0 (x-y;\kappa )f(y),
$$
where the kernel $G_0 (x;\kappa )$ is a diagonal matrix
$$
G_0 (x;\kappa )=\mathrm{diag} [r_{\leftarrow} (x;\kappa ),  r_{\rightarrow} (x;\kappa ),r_{\downarrow} (x;\kappa ),r_{\uparrow} (x;\kappa ) ].
$$
Passing through the Fourier transformation, we can see easily the following lemma.

\begin{lemma}
Let $\delta=\{ \delta_{x0} \}_{x\in {\bf Z}^2}$ where $\delta_{xy}$ is the Kronecker delta for $x,y\in {\bf Z}^2$.
The kernel $G_0 (x;\kappa )$ is the fundamental solution to the equation (\ref{S3_eq_helmholtz}) in the sense
\begin{gather*}
(U_0 -e^{i\kappa} )G_0 (\cdot ;\kappa )=\mathrm{diag} [\delta , \delta , \delta , \delta ] ,
\end{gather*}
for $\kappa \in {\bf C} \setminus {\bf R}$.
In particular, we have $(R_0 (\kappa )f)(x)=\sum_{y\in {\bf Z}^2} G_0 (x-y;\kappa )f(y)$ for $f\in \mathcal{B}$.
\label{S3_lem_green1}
\end{lemma}

Let us compute $G_0 (x;\kappa )$ explicitly as follows.
We have 
$$
r_{\leftarrow} (x;\kappa )= \frac{1}{(2\pi )^2} \int _{{\bf T}^2} \frac{e^{i\langle x,\xi\rangle}}{e^{i\xi_1} -e^{i\kappa}} d\xi = \frac{1}{(2\pi )^2} \int _{{\bf T}} \frac{e^{ix_1 \xi_1}}{e^{i\xi_1} -e^{i\kappa}} d\xi_1 \int _{{\bf T}} e^{ix_2 \xi_2} d\xi_2 .
$$
Since we have $\int _{{\bf T}} e^{ix_2 \xi_2 } d\xi_2 = 2\pi \delta _{x_2 0} $, we obtain 
\begin{equation}
r_{\leftarrow} (x;\kappa )= \frac{\delta_{x_2 0}}{2\pi} \int _{{\bf T}} \frac{e^{ix_1 \xi_1}}{e^{i\xi_1} - e^{i\kappa}} d\xi_1 .
\label{S3_eq_Green1}
\end{equation}
By the similar argument, we also have 
\begin{gather}
r_{\rightarrow} (x;\kappa )= \frac{\delta_{x_2 0}}{2\pi} \int _{{\bf T}} \frac{e^{ix_1 \xi_1}}{e^{-i\xi_1} - e^{i\kappa}} d\xi_1 ,
\label{S3_eq_Green2} \\
r_{\downarrow} (x;\kappa )= \frac{\delta_{x_1 0}}{2\pi} \int _{{\bf T}} \frac{e^{ix_2 \xi_2}}{e^{i\xi_2} - e^{i\kappa}} d\xi_2 ,
\label{S3_eq_Green3} \\
r_{\uparrow} (x;\kappa )= \frac{\delta_{x_2 0}}{2\pi} \int _{{\bf T}} \frac{e^{ix_2 \xi_2}}{e^{-i\xi_2} - e^{i\kappa}} d\xi_2 .
\label{S3_eq_Green4}
\end{gather}
Then we can apply \cite[Lemma 2.5]{KKMS} in order to prove the following formulas.

\begin{lemma}
Let $ F(s) $ be the characteristic function of the set $\{ s\in {\bf R} \ ; \ s\geq 0 \}  $. 
We put $ \kappa_{\pm} = \theta -i\log (1 \mp \epsilon )$ for $\epsilon >0$ and $ \theta \in [0,2\pi )$.
We have 
\begin{gather}
r _{\leftarrow} (x;\kappa _+ )= \delta_{x_2 0} F(x_1 -1) e^{i\kappa_+ (x_1 -1)} , \label{S3_eq_greenexplicit1+} 
\\
r _{\rightarrow} (x;\kappa _+ )= \delta_{x_2 0} F(-x_1 -1) e^{-i\kappa_+ (x_1 +1)} , \label{S3_eq_greenexplicit2+} 
\\
r _{\downarrow} (x;\kappa _+ )= \delta_{x_1 0} F(x_2 -1) e^{i\kappa_+ (x_2 -1)} , \label{S3_eq_greenexplicit3+} 
\\
r _{\uparrow} (x;\kappa _+ )= \delta_{x_1 0} F(-x_2 -1) e^{-i\kappa_+ (x_2 +1)} , \label{S3_eq_greenexplicit4+}
\end{gather}
and
\begin{gather}
r _{\leftarrow} (x;\kappa _- )= -\delta_{x_2 0} F(-x_1 ) e^{i\kappa_- (x_1 -1)} , \label{S3_eq_greenexplicit1-} \\
r _{\rightarrow} (x;\kappa _- )=- \delta_{x_2 0} F(x_1 ) e^{-i\kappa_- (x_1 +1)} , \label{S3_eq_greenexplicit2-} \\
r _{\downarrow} (x;\kappa _- )= -\delta_{x_1 0} F(-x_2 ) e^{i\kappa_- (x_2 -1)} , \label{S3_eq_greenexplicit3-} \\
r _{\uparrow} (x;\kappa _- )= -\delta_{x_1 0} F(x_2 ) e^{-i\kappa_- (x_2 +1)} . \label{S3_eq_greenexplicit4-}
\end{gather}
\label{S3_lem_green2}
\end{lemma}

Even if we take limits of (\ref{S3_eq_greenexplicit1+})-(\ref{S3_eq_greenexplicit4-}) as $\epsilon \downarrow 0 $ in the weak $*$ sense, the formulas (\ref{S3_eq_greenexplicit1+})-(\ref{S3_eq_greenexplicit4-}) hold.
Letting ${\bf e} _{\leftarrow} = [1,0,0,0]^{\mathsf{T}} $, $ {\bf e} _{\rightarrow} = [0,1,0,0]^{\mathsf{T}} $, ${\bf e} _{\downarrow} = [0,0,1,0]^{\mathsf{T}} $, $ {\bf e} _{\uparrow} = [0,0,0,1]^{\mathsf{T}} $, $ R_0 (\theta +i0 )f$ for $f\in\mathcal{B}$ can be represented by 
\begin{gather}
\begin{split}
& (R_0 (\theta +i0 )f)(x) \\
&= e^{i \theta (x_1 -1)} \sum_{y_1 \leq x_1 -1 } e^{-i\theta y_1}  f_{\leftarrow} ( y_1 ,x_2 ){\bf e} _{\leftarrow} + e^{-i\theta (x_1 +1)} \sum _{y_1 \geq x_1 +1} e^{i\theta y_1} f_{\rightarrow} (y_1 , x_2 ) {\bf e} _{\rightarrow} \\
&\quad +e^{i \theta (x_2 -1)} \sum_{y_2 \leq x_2 -1 } e^{-i\theta y_2}  f_{\downarrow} ( x_1 ,y_2 ){\bf e} _{\downarrow} + e^{-i\theta (x_2 +1)} \sum _{y_2 \geq x_2 +1} e^{i\theta y_2} f_{\uparrow} (x_1 , y_2 ) {\bf e} _{\uparrow}  ,
\end{split}
\label{S3_eq_resolvent0+}
\end{gather} 
and
$ R_0 (\theta -i0 )f$ can be represented by 
\begin{gather}
\begin{split}
& (R_0 (\theta -i0 )f)(x) \\
&=- e^{i \theta (x_1 -1)} \sum_{y_1 \geq x_1  } e^{-i\theta y_1}  f_{\leftarrow} ( y_1 ,x_2 ){\bf e} _{\leftarrow} - e^{-i\theta (x_1 +1)} \sum _{y_1 \leq x_1 } e^{i\theta y_1} f_{\rightarrow} (y_1 , x_2 ) {\bf e} _{\rightarrow} \\
&\quad -e^{i \theta (x_2 -1)} \sum_{y_2 \geq x_2  } e^{-i\theta y_2}  f_{\downarrow} ( x_1 ,y_2 ){\bf e} _{\downarrow} - e^{-i\theta (x_2 +1)} \sum _{y_2 \leq x_2 } e^{i\theta y_2} f_{\uparrow} (x_1 , y_2 ) {\bf e} _{\uparrow}  .
\end{split}
\label{S3_eq_resolvent0-}
\end{gather} 
Note that the summations on the right-hand side of (\ref{S3_eq_resolvent0+})-(\ref{S3_eq_resolvent0-}) converge due to Lemma \ref{S2_lem_l1prop}.
Then we obtain the limiting absorption principle for $ R_0 (\kappa )$.

\begin{lemma}
Let $J $ be an arbitrary compact interval in $[0,2\pi )$.
\begin{enumerate}
\item For $ \theta \in [0,2\pi )$, we have $ R_0 (\theta \pm i0 ) \in {\bf B} ( \mathcal{B} ; \mathcal{B}^* )$ in the weak $*$ topology.
In particular, there exists a constant $c>0$ such that $\| R_0 (\theta \pm i0 )f \| _{\mathcal{B}^*} \leq c\| f\| _{\mathcal{B}} $ where $\theta$ varies on $J$.

\item 
The mapping $J\ni \theta \mapsto (R_0 (\theta \pm i0 )f,g)$ for $f,g \in \mathcal{B}$ is continuous. 

\end{enumerate}
\label{S3_lem_LAP0}
\end{lemma}

We can see the asymptotic behavior of $ R_0 (\theta \pm i0 )f$ at infinity from formulas (\ref{S3_eq_resolvent0+})-(\ref{S3_eq_resolvent0-}).

\begin{lemma}
Let $f\in \mathcal{B}$.
We have
\begin{gather}
\begin{split}
R_0 ( \theta +i0 )f \simeq & \, F(x_1 ) e^{i\theta (x_1 -1)} \sum _{y_1 \in {\bf Z}} e^{-i\theta y_1} f_{\leftarrow} (y_1 ,x_2 ) {\bf e}_{\leftarrow} \\
& \, + F(-x_1 ) e^{-i \theta (x_1 +1 )} \sum _{y_1 \in {\bf Z}} e^{i \theta y_1} f_{\rightarrow} (y_1, x_2 ) {\bf e}_{\rightarrow} \\
& \, +F(x_2 ) e^{i\theta (x_2 -1)} \sum _{y_2 \in {\bf Z}} e^{-i\theta y_2} f_{\downarrow} (x_1 , y_2 ) {\bf e}_{\downarrow}  \\
& \, + F(-x_2 ) e^{-i \theta (x_2 +1 )} \sum _{y_2 \in {\bf Z}} e^{i \theta y_2} f_{\uparrow} (x_1, y_2 ) {\bf e}_{\uparrow} ,
\end{split}
\label{S3_eq_asymptotic0+}
\end{gather}
and
\begin{gather}
\begin{split}
R_0 ( \theta -i0 )f \simeq & \, - F(-x_1 ) e^{i\theta (x_1 -1)} \sum _{y_1 \in {\bf Z}} e^{-i\theta y_1} f_{\leftarrow} (y_1 ,x_2 ) {\bf e}_{\leftarrow} \\
& \, - F(x_1 ) e^{-i \theta (x_1 +1 )} \sum _{y_1 \in {\bf Z}} e^{i \theta y_1} f_{\rightarrow} (y_1, x_2 ) {\bf e}_{\rightarrow} \\
& \, -F(-x_2 ) e^{i\theta (x_2 -1)} \sum _{y_2 \in {\bf Z}} e^{-i\theta y_2} f_{\downarrow} (x_1 , y_2 ) {\bf e}_{\downarrow}  \\
& \, - F(x_2 ) e^{-i \theta (x_2 +1 )} \sum _{y_2 \in {\bf Z}} e^{i \theta y_2} f_{\uparrow} (x_1, y_2 ) {\bf e}_{\uparrow} .
\end{split}
\label{S3_eq_asymptotic0-}
\end{gather}
\label{S3_lem_asymptotic0}
\end{lemma}

Proof.
Take $f\in \mathcal{B} $ such that $ \mathrm{supp} f$ is finite.
Now we denote the right-hand side of (\ref{S3_eq_asymptotic0+}) by $ I _{1,+} (x)+ I_{1,-} (x)+ I_{2,+} (x)+ I_{2,-} (x)$.
We have 
\begin{gather*}
\begin{split}
&( R_0 (\theta +i0)f)(x)-(I _{1,+} (x)+ I_{1,-} (x)+ I_{2,+} (x)+ I_{2,-} (x)) \\
=& \, e^{i\theta (x_1 -1)} \sum_{y_1 \in {\bf Z}} e^{-i\theta y_1} (F(x_1 -y_1 -1)-F(x_1)) f_{\leftarrow} (y_1 ,x_2 ) {\bf e} _{\leftarrow} \\
& \, + e^{-i\theta (x_1 +1)} \sum _{y_1 \in {\bf Z}} e^{i\theta y_1} (F(-x_1 +y_1 -1 )-F(-x_1)) f_{\rightarrow} (y_1 ,x_2) {\bf e} _{\rightarrow} \\
& \, +e^{i\theta (x_2 -1)} \sum_{y_2 \in {\bf Z}} e^{-i\theta y_2} (F(x_2 -y_2 -1)-F(x_2)) f_{\downarrow} (x_1 ,y_2 ) {\bf e} _{\downarrow} \\
& \, + e^{-i\theta (x_2 +1)} \sum _{y_2 \in {\bf Z}} e^{i\theta y_2} (F(-x_2 +y_2 -1 )-F(-x_2)) f_{\uparrow} (x_1 ,y_2) {\bf e} _{\uparrow}  .
\end{split}
\end{gather*}
Thus $ ( R_0 (\theta +i0)f)(x)-(I _{1,+} (x)+ I_{1,-} (x)+ I_{2,+} (x)+ I_{2,-} (x)) $ vanishes except for a finite number of $x\in {\bf Z}^2 $.
This implies the formula (\ref{S3_eq_asymptotic0+}).
For any $f\in \mathcal{B}$ and $\epsilon >0$, there exists $g\in \mathcal{B}$ such that $g$ has a finite support and satisfies $ \| f-g\| _{\mathcal{B}} < \epsilon $.
Then the formula (\ref{S3_eq_asymptotic0+}) holds for any $f\in \mathcal{B}$ due to Lemma \ref{S3_lem_LAP0}.
The proof of (\ref{S3_eq_asymptotic0-}) is similar.
\qed

\subsection{Multi-dimensional unique continuation for QW}

In this subsection, we consider the unique continuation property for generalized eigenfunctions of $d$DQWs as follows.

\begin{lemma}
Suppose that a ${\bf C}^{2d}$-valued sequence $\psi$ on ${\bf Z}^d$ satisfies the equation $(U-e^{i\theta} )\psi =0$.
For any $x\in {\bf Z}^d$, the value $\psi (x)$ is determined uniquely by $\psi (x+e_j )$, $j=1,\ldots ,d$.
\label{S3_lem_UCP}
\end{lemma}

Proof.
The unique continuation property for 1DQW is well-known since $(U-e^{i\theta} )\psi =0$ on ${\bf Z}$ can be reduced to a first order recurrence formula.
Let us prove the lemma for $d\geq 2$.
We put 
$$
C(x)=( \vec{c}_j (x))_{j=1}^{2d}, \quad x\in {\bf Z}^d ,
$$
where $\vec{c}_j (x)$ are row vectors of $C(x)$.
In the proof of this lemma, we use the similar notation for row vectors of some matrices.
We also define matrices $A^{(k)}_{\theta} (x)=( \vec{a}^{(k)}_{j,\theta} (x)) _{j=1}^{2d} $, $j=1,\ldots ,d$, and $M_{\theta} (x)=(\vec{m} _{j,\theta} (x))_{j=1}^{2d} $ for $x\in {\bf Z}^d$ by
\begin{gather*}
\vec{a}^{(k)}_{j,\theta} (x) =\left\{ 
\begin{split}
  \vec{c}_j (x) &, \quad j=2k-1 , \\
 e^{i\theta} (e_j)^{\mathsf{T}} &, \quad j=2k, \\
0 &, \quad \text{otherwise} ,
\end{split}
\right.
\quad
\vec{m}_{j,\theta} (x) =\left\{ 
\begin{split}
e^{i\theta} (e_j)^{\mathsf{T}} &, \quad j \text{ is odd} , \\
  \vec{c}_j (x) &, \quad j \text{ is even},
\end{split}
\right.
\end{gather*}
noting that $e_j$, $j=1,\ldots,2d$, are the standard basis of ${\bf R}^{2d}$.
The equation $(U-e^{i\theta} )\psi =0 $ on ${\bf Z}^d$ can be rewritten as 
$$
\sum_{k=1}^d A^{(k)}_{\theta} (x+e_j ) \psi (x+e_k )= M_{\theta} (x) \psi (x), \quad x\in {\bf Z}^d .
$$
In view of the assumption for $C(x)$, we have 
$$
\det M_{\theta} (x)= e^{di\theta} \det ( c_{2l_1 ,2l_2} (x)) _{l_1 ,l_2 =1}^d \not= 0.
$$
Thus we obtain 
$$
\psi (x)=M_{\theta} (x)^{-1} \sum_{k=1}^d A^{(k)}_{\theta} (x+e_j ) \psi (x+e_k )   ,
$$
which implies the lemma.
\qed

\medskip

As a consequence of this lemma, we obtain the following assertion.

\begin{cor}
Let $O_N (x) =\{ y\in {\bf Z}^d \setminus \{ x\} \ ; \ y_j -x_j \geq 0 , j=1,\ldots ,d , \sum _{j=1}^d (y_j -x_j ) =N \}$ for $x\in {\bf Z}^d$ and a positive integer $N$.   
If the solution $\psi$ to the equation $(U-e^{i\theta} )\psi =0$ is known in the subset $O_N (x)$, the value $\psi (x)$ is determined uniquely from $\psi \big| _{O_N (x)}$.
In particular, if $\psi =0$ in $O_N (x)$, we have $\psi (x)=0$.
\label{S3_cor_UCP2}
\end{cor}

\subsection{Radiation condition}
Lemma \ref{S3_lem_asymptotic0} implies the radiation condition which guarantees the uniqueness of solutions to the equation $(U-e^{i\theta} )u=f$ for $f\in \mathcal{B}$.
In view of the anisotropy of the asymptotic behavior of $R_0 (\theta \pm i0 )f$, we define the radiation condition  as follows.

Let us introduce the operator
$$
( B_{\pm} f )(x)= [ F( \pm x_1 ) f_{\leftarrow} (x) , F(\mp x_1 ) f_{\rightarrow} (x), F(\pm x_2 ) f_{\downarrow} (x) , F(\mp x_2 ) f_{\uparrow} (x)] ^{\mathsf{T}},
$$
for $x\in {\bf Z}^2 $.

\begin{definition}
The solutions $u^{(\pm )} \in \mathcal{B}^* $ to the equation $(U-e^{i\theta} )u^{(\pm )} =f$ for $f\in \mathcal{B} $ are incoming (for $+$) or outgoing (for $-$) if $u^{(\pm )} $ satisfy
\begin{equation}
B_{\pm} S u^{(\pm )} -e^{i\theta} u^{(\pm )} \in \mathcal{B}_0^* .
\label{S3_eq_radcondi}
\end{equation}
\label{S3_def_radiationcondition}
\end{definition}

\begin{lemma}
Suppose that $ u^{(\pm )} \in \mathcal{B}^* $ satisfy the equation $ (U-e^{i\theta} )u^{(\pm  )} =0$ and the condition (\ref{S3_eq_radcondi}).
Then we have $u^{(\pm)} =0$.
\label{S3_lem_uniqueness}
\end{lemma} 

Proof.
We prove for $ u^{(+)} $ and the proof for $ u^{(-)} $ is similar.
Due to the equation $ (U-e^{i\theta} )u^{(+)} =0$, $u^{(+)} _{\leftarrow} $ satisfies $u^{(+)} _{\leftarrow} (x+e_1 ) =e^{i\theta} u _{\leftarrow}^{(+)} (x)$ for $x \in {\bf Z}^2 \setminus D$ with $x_1 \leq -n_0 -2 $ or $|x_2| \geq n_0 +1 $.
In particular, we have $| u_{\leftarrow} ^{(+)} (x+e_1 )| = |u_{\leftarrow} ^{(+)} (x ) |$.
This equality and the radiation condition (\ref{S3_eq_radcondi}) imply $ \liminf _{x_1 \to -\infty} | u_{\leftarrow} (x_1 ,x_2 ) | =0$ for every $x_2$. 
Then we have
\begin{equation}
u^{(+)} _{\leftarrow} (x)=0 \quad \text{in} \quad D_{\leftarrow}^+ =\{ x\in {\bf Z}^2 \ ; \ x_1 \leq -n_0 -1 \ \text{or} \ |x_2| \geq n_0 +1 \} .
\label{S3_eq_vanish1}
\end{equation}
By the similar way, we have from the equation $(U-e^{i\theta} )u^{(+)} =0$ and the radiation condition (\ref{S3_eq_radcondi})
\begin{gather}
u^{(+)} _{\rightarrow} (x)=0 \quad \text{in} \quad D_{\rightarrow}^+ =\{ x\in {\bf Z}^2 \ ; \ x_1 \geq n_0 +1 \ \text{or} \ |x_2| \geq n_0 +1 \} ,
\label{S3_eq_vanish2} \\
u^{(+)}_{\downarrow} (x)=0 \quad \text{in} \quad D_{\downarrow}^+ =\{ x\in {\bf Z}^2 \ ; \ x_2 \leq -n_0 -1 \ \text{or} \ |x_1| \geq n_0 +1 \} ,
\label{S3_eq_vanish3} \\
u^{(+)} _{\uparrow} (x)=0 \quad \text{in} \quad D_{\uparrow}^+ =\{ x\in {\bf Z}^2 \ ; \ x_2 \geq n_0 +1 \ \text{or} \ |x_1| \geq n_0 +1 \} .
\label{S3_eq_vanish4}
\end{gather}

We put 
\begin{gather*}
\partial D = \{ x\in {\bf Z}^2 \setminus D \ ; \ \exists y\in D \ \text{such that} \ |x-y|=1 \} , \\
E_{\leftarrow}^+ = \{ x\in {\bf Z}^2 \ ; \ x_1 = n_0 +1, n_0 +2 , \ |x_2| \leq n_0 \} , \\
E_{\rightarrow}^+ = \{ x\in {\bf Z}^2 \ ; \ x_1 = -n_0 -2, -n_0 -1, \ |x_2|\leq n_0 \} , \\
E_{\downarrow}^+ = \{ x\in {\bf Z}^2 \ ; \ |x_1| \leq n_0 , \ x_2 = n_0 +1, n_0 +2 \} , \\
E_{\uparrow}^+ = \{ x\in {\bf Z}^2 \ ; \ |x_1| \leq n_0 , \ x_2 = -n_0 -2, -n_0 -1 \} .
\end{gather*}
By (\ref{S3_eq_vanish1})-(\ref{S3_eq_vanish4}) and the definition of the operator $U=SC$, we have
\begin{equation}
\sum_{x\in D\cup \partial D} | (Uu^{(+)} )(x) |^2 = \sum_{x\in D} | C(x) u^{(+)} (x)|^2 + \sum_{p\in \{ \leftarrow, \rightarrow, \downarrow, \uparrow \}} \sum _{x\in E_{p}^+ } | u^{(+)}_{p} (x) |^2
\label{S3_eq_inout}
\end{equation}
Since $C(x)$ is unitary, we have $|C(x) u^{(+)} (x)| ^2 = | u^{(+)} (x)|^2 $ for every $x\in D$.
Due to the equation $ (U-e^{i\theta} )u^{(+)} =0$, we also have $ \sum_{x\in D \cup \partial D} | (Uu^{(+)} )(x)|^2 = \sum_{x\in D\cup \partial D} | u^{(+)} (x)|^2$.
Thus (\ref{S3_eq_inout}) can be rewritten as 
$$
\sum_{p\in \{ \leftarrow, \rightarrow, \downarrow, \uparrow \}}  \sum _{x\in E_{p}^+ \setminus \partial D} | u^{(+)}_{p} (x) |^2 =0
$$
This implies $u_{\leftarrow}^{(+)} (n_0 +2,x_2)=u_{\rightarrow}^{(+)} (-n_0 -2,x_2 )=0$ for $|x_2| \leq n_0$, and $u_{\downarrow}^{(+)} (x_1 ,n_0 +2)=u_{\uparrow}^{(+)} (x_1 , -n_0 -2 )=0$ for $|x_1| \leq n_0$.
By using the equation $(U-e^{i\theta} )u^{(+)} =0$ in the region ${\bf Z}^2 \setminus D$, we obtain $u^{(+)} =0$ in ${\bf Z}^2 \setminus D$.
Due to Corollary \ref{S3_cor_UCP2}, $u^{(+)}=0 $ is extended in $D$.
\qed

\medskip

As a direct consequence of this lemma, we can show immediately the uniqueness of the incoming (for $+$) or outgoing (for $-$) solution to the equation $ (U_0 -e^{i\theta} )u=f$ for $ f\in \mathcal{B}$.

\begin{cor}
The solution $u^{(\pm )} \in \mathcal{B}^*$ to the equation $ (U_0 -e^{i\theta} )u^{(\pm )} =f$ for $f\in \mathcal{B} $ is incoming (for $+$) or outgoing (for $-$) if and only if $u^{(\pm )}=R_0 (\theta \pm i0 )f$.  
\label{S3_cor_helmhltzunique}
\end{cor}

Proof.
$R_0 ( \theta \pm i0 )f$ obviously satisfy the radiation condition (\ref{S3_eq_radcondi}) in view of Lemma \ref{S3_lem_asymptotic0}.
For the proof of uniqueness, we assume that the solutions $v^{(\pm)} \in \mathcal{B}^*$ to the equation $(U_0 -e^{i\theta})v^{(\pm )} =f$ are incoming (for $+$) or outgoing (for $-$).
Thus $w^{(\pm )} =R_0 (\theta \pm i0 )f -v^{(\pm)}$ satisfy the equation $(U_0 -e^{i\theta} )w^{(\pm)} =0$ and the condition (\ref{S3_eq_radcondi}).
Lemma \ref{S3_lem_uniqueness} implies $w^{(\pm )} =0$.
\qed

\medskip

We also have the absence of eigenvalues of $U$.
\begin{cor}
We have $\sigma_p (U)= \emptyset $.
\label{S3_cor_absenceEV}
\end{cor}

Proof.
Recall that $\sigma_{ess} (U)= \{ e^{i\theta} \ ; \ \theta \in [0,2\pi ) \}$.
Then $\sigma_p (U)\subset \sigma_{ess} (U)$.
If there exists an eigenvalue $e^{i\theta}$, the corresponding eigenfunction $\psi \in \ell^2$ satisfies the equation $(U-e^{i\theta}) \psi =0$ and the condition (\ref{S3_eq_radcondi}).
Lemma \ref{S3_lem_uniqueness} implies $\psi =0$.
This is a contradiction.
\qed


\section{Generalized eigenfunction}
\subsection{Combinatorial construction of generalized eigenfunction}
Before we derive the spectral theory for $U$, we mention a combinatorial construction of generalized eigenfunctions.
This construction is based on a long time behavior of a dynamics of the QW.
Our argument in this subsection is an analogue of \cite{KKMS} which is a special case of \cite[Theorem 3.1]{HiSe}.

Let $ \chi : \mathcal{B}^* \to \ell^2 (D;{\bf C}^4 )$ be defined by $(\chi u)(x)=u(x)$ for $x\in D$. 
We also introduce the operator $\chi^* : \ell^2 (D;{\bf C}^4 )\to \mathcal{B}^* $ by $(\chi^* \psi )(x)=\psi (x)$ for $x\in D$ and $(\chi^* \psi )(x)=0$ for $x\in {\bf Z}^2 \setminus D$.
We define the $4(2n_0 +1)^2 \times 4(2n_0 +1)^2$ submatrix 
$$
U_D = \chi U \chi^* .
$$
Fixing a basis of $\ell^2 (D;{\bf C}^4) = {\bf C} ^{4(2n_0 +1)^2}$, we can obtain an explicit representation of the matrix $U_D $. 
In this paper, we omit it.

\begin{lemma}
Eigenvalues of $U_D$ lie in the subset $\{ \lambda \in {\bf C} \ ; \ |\lambda| <1\} $.
\label{S4_lem_EV_En}
\end{lemma}

Proof.
For an eigenvector $\psi \in \ell^2 (D;{\bf C}^4 )$ associated with an eigenvalue $\lambda$, we have 
$$
|\lambda |^2 \| \psi \|^2 _{\ell^2 (D;{\bf C}^4 ) } = \| U_D \psi \|^2 _{\ell^2 (D;{\bf C}^4 ) } \leq \| U \chi^* \psi \|^2  _{\ell^2 ({\bf Z}^2 ;{\bf C}^4 )} = \| \psi \|^2 _{\ell^2 (D;{\bf C}^4 )} .
$$
This implies $|\lambda  | \leq 1$.
Now we suppose $ |\lambda | =1$.
This implies $\| U_D \psi \|^2 _{\ell^2 (D;{\bf C}^4 )} = \| U\chi^* \psi \|^2 _{\ell^2 ({\bf Z}^2 ;{\bf C}^4 )}$.
In view of the definition of $\chi$ and $\chi^*$, we have $(1-\chi^* \chi )U\chi^* \psi =0$.
Thus we obtain 
$$
U\chi^* \psi = \chi^* \chi U\chi^* \psi =\lambda \chi^* \psi ,
$$
and $\chi^* \psi $ is an eigenfunction of $U$ with a finite support.
It follows $\chi^* \psi =0$ from Corollary \ref{S3_cor_absenceEV}.
This is a contradiction.
\qed

\medskip

Lemma \ref{S4_lem_EV_En} implies the convergence of the series 
$$
\sum _{m=0}^{\infty} e^{-im\theta} U_D^m , \quad \theta \in [0,2\pi ) ,
$$
in view of the Jordan canonical form of $U_D$.
Namely, there exists a $4(2n_0 +1)^2 \times 4(2n_0 +1)^2$ regular matrix $G$ such that 
$$
G ^{-1} U_D G = J (\lambda _1 ,k_1 )\oplus \cdots \oplus J(\lambda_r ,k_r ),
$$
for a positive integer $r$ where $J(\lambda_j ,k_j)$ is the Jordan block for an eigenvalue $\lambda_j$ with algebraic multiplicity $k_j$.

Now let us derive a construction of a generalized eigenfunction of $U$ with a single incident wave along an incoming path.
Note that a generalized eigenfunction of $U$ with multiple incident waves can be written by a linear combination of the generalized eigenfunctions associated with a single incident wave. 

\begin{lemma}
Take an integer $b\in [-n_0 ,n_0]$.
Let $\Psi_0 \in \mathcal{B}^* $ be given by 
\begin{gather*}
\Psi _0 (x)= \left\{ 
\begin{split}
e^{i\theta x_1} {\bf e} _{\leftarrow} &, \quad x=(x_1 , b), \quad x_1 \geq n_0 +1, \\
0 &, \quad \text{otherwise} ,
\end{split}
\right.
\end{gather*}
for $\theta \in [0,2\pi )$.
We define $\Psi_t$ for every positive integer $t$ by $\Psi_t = U \Psi_{t-1} $.
Then there exists a limint
$$
\Psi_{\infty} (x):= \lim_{t\to \infty} e^{-it\theta } \Psi _t (x) ,\quad x\in {\bf Z}^2 ,
$$
and $\Psi_{\infty}\in \mathcal{B}^*$ satisfies $U\Psi_{\infty} =e^{i\theta} \Psi_{\infty}$ on ${\bf Z}^2 $. 

\label{S4_lem_gefconstruction}
\end{lemma}

Proof.
The initial state $\Psi_0$ is an incoming flow on $(x_1 ,b)$ for $x_1 \geq n_0 +1$.
The outgoing flow occurs due to the perturbation in the region $D$. 
Once the flow comes out of $D$, it goes away. 
In view of this dynamics, we split $\Psi_t$ into three parts 
$$
\Psi_t = \chi^* \chi \Psi_t + (1-\chi^* \chi )(\Psi_t - e^{it\theta} \Psi_0 ) +e^{it\theta} (1-\chi^* \chi ) \Psi_0 .
$$
The first term on the right-hand side is the state in $D$.
The second and third term on the right-hand side are the outgoing flow and the incoming flow on ${\bf Z}^2 \setminus D$, respectively.

Letting $\phi_t = \chi \Psi_t$, we have 
$$
\phi_0 =0 , \quad \phi_{t+1} = U_D \phi_t + e^{it\theta} \chi U\Psi_0 , \quad t\geq 0,
$$
where we have used the equality $ \chi U(1-\chi^* \chi)U^t \Psi_0 = e^{it\theta} \chi U\Psi_0$.
The solution of this recurrence formula is 
$$
\phi_t = e^{i\theta (t-1)} \sum _{m=0}^{t-1} e^{-im\theta} U_D^m \chi U \Psi_0 , \quad t\geq 0.
$$
In view of Lemma \ref{S4_lem_EV_En}, the limit
$$
\phi_{\infty} := \lim_{t\to \infty} e^{-it\theta} \phi_t = e^{-i\theta} (1-e^{-i\theta} E)^{-1} \chi U \Psi_0 
$$
exists and obtain the equation 
\begin{equation}
U_D \phi _{\infty} + \chi U \Psi_0 = e^{i\theta} \phi _{\infty} .
\label{S4_eq_inteq}
\end{equation}

Let us derive the outgoing flow.
We define the subset 
$$
B=\{ x\in D \ ; \ | x_1 |= n_0 \ \text{or} \ |x_2| =  n_0 \} .
$$
Let $\delta_B$ be the characteristic function of the subset $B$.
The outgoing flow for $t\geq 2 $ is given by
$$
(1-\chi^* \chi ) (\Psi_t - e^{it\theta} \Psi_0 ) = (1-\chi^* \chi ) \sum _{m=1}^{t-1} U^m f _{t-m} ,
$$
where the source $f_{t-m}$ is defined by $f_{t-m} =\delta _B \chi^* \phi _{t-m} $.
For $t\geq 2$, the value of the outgoing flow $\Psi _{t,out} :=(1-\chi^* \chi ) (\Psi_t - e^{it\theta} \Psi_0 )$ at every point $x\in {\bf Z}^2 \setminus D$ is 
\begin{gather*}
\Psi_{t,out} (x_1,x_2) =(U ^{\mu_{x_1} } f_{t-\mu_{x_1}})(x_1,x_2) \quad \text{for} \quad |x_1| >n_0, \ |x_2|\leq n_0,\\
\Psi_{t,out} (x_1,x_2)=( U ^{\mu_{x_2} } f_{t-\mu_{x_2}})(x_1 ,x_2) \quad  \text{for} \quad |x_1|\leq n_0, \ |x_2|>n_0 +1,
\end{gather*}
or $\Psi _{t,out} (x)=0$ for other $x\in {\bf Z}^2 \setminus D$ where  
\begin{gather*}
\mu_{x_1 } = \left\{ \begin{split}
 x_1 -n_0 , & \quad x_1 \geq n_0 +1, \\
 -x_1 -n_0 ,&\quad x_1 \leq -n_0 -1, 
\end{split} 
\right.
\quad \mu_{x_2} = \left\{ \begin{split} 
 x_2 -n_0 ,&\quad x_2 \geq n_0 +1, \\
- x_2 -n_0 ,&\quad x_2\leq -n_0-1.
\end{split}
\right.
\end{gather*}
Then the limit $ \Psi _{out} := \lim_{t\to \infty} e^{-it\theta}  (1-\chi^* \chi ) (\Psi_t - e^{it\theta} \Psi_0 ) $ is given by 
$$
\Psi _{out} (x) =\lim_{t\to \infty} e^{-it \theta} ( U^{\mu_x} f_{t-\mu_x} )(x)= e^{-i\mu_x \theta} (Uf_{\infty} )(\widetilde{x}) ,
$$
for $f_{\infty} = \delta_B \chi ^* \phi _{\infty} $, $\mu_x = \mu_{x_1} $ or $\mu_{x_2} $, and $\widetilde{x} = (\pm ( n_0+1) ,x_2 )$ or $(x_1 , \pm (n_0 +1))$.

Finally, we show the equation $U\Psi _{\infty} =e^{i\theta} \Psi _{\infty} $.
Note that 
$$
\Psi _{\infty} = \chi^* \phi _{\infty} + \Psi _{out} + \Psi_0 \quad \text{on} \quad {\bf Z}^2 .
$$
Then we have
\begin{equation}
\chi^* \chi U \Psi _{\infty} = \chi^* U_D \phi_{\infty} +\chi^* \chi U \Psi_0 + \chi^* \chi U\Psi _{out} = e^{i\theta} \chi^* \phi _{\infty} ,
\label{S4_eq_eq11}
\end{equation}
in view of (\ref{S4_eq_inteq}) and $\chi U\Psi _{out} =0$.
We also have
\begin{gather}
\begin{split}
(1-\chi^* \chi ) U\Psi _{\infty} &= (1-\chi^* \chi )U( \chi^* \phi _{\infty} + \Psi_0 ) + U\Psi _{out} \\ 
&= (1-\chi^* \chi ) \delta _{\partial D} Uf_{\infty} +e^{i\theta} \Psi_0 + e^{i\theta} \Psi_{out} -\delta _{\partial D} Uf_{\infty} \\
&= e^{i\theta} (\Psi _{out} +\Psi _0 ),
\end{split}
\label{S4_eq_eq12}
\end{gather}
where $\delta_{\partial D}$ is the characteristic function of the subset $\partial D$ (see the proof of Lemma \ref{S3_lem_uniqueness} for the definition of $\partial D$).
Plugging (\ref{S4_eq_eq11}) and (\ref{S4_eq_eq12}), we obtain the equation $U\Psi _{\infty} =e^{i\theta} \Psi _{\infty} $.
\qed

\medskip

\textit{Remark.}
For 1DQWs, a combinatorial formula of the S-matrix was derived by \cite{KKMS} which is based on counting paths of quantum walkers in an interval on which $U\not= U_0 $.
For multi-dimensional cases, this approach is more complicated.
However, as has been derived in Lemma \ref{S4_eq_inteq}, we can construct generalized eigenfunctions as a long time limit of a dynamics of QWs.

\subsection{Limiting absorption principle for position-dependent QW}

We have proved the existence and a construction of generalized eigenfunctions in $\mathcal{B}^*$.
In the following, we discuss the spectral theory for the operator $U$ in order to characterize rigorously the set of generalized eigenfunctions in $\mathcal{B}^*$.
By using the distorted Fourier transformation associated with $U$, we can see that the S-matrix appears in the outgoing scattered wave $\Psi _{out}$.

Here we prove the existence of the limits $R(\theta \pm i0 ) = \lim _{\epsilon \downarrow 0} R(\theta -i\log (1\mp \epsilon ))$ in ${\bf B} (\mathcal{B};\mathcal{B}^* )$.
In the following argument, we put 
\begin{equation}
V=U-U_0 . 
\label{S4_eq_V}
\end{equation}
The well-known resolvent equations hold :
\begin{equation}
R(\kappa )= R_0 (\kappa )- R_0 (\kappa )VR(\kappa )= R_0 (\kappa )-R(\kappa )VR_0 (\kappa ),
\label{S4_eq_resolventeq}
\end{equation}
for $ \kappa \in {\bf C} \setminus {\bf R} $.
In fact, these equalities follow from 
$$
(U_0 -e^{i\kappa} ) R(\kappa )= 1-VR (\kappa ) , \quad (U-e^{i\kappa} ) R_0 (\kappa )= 1+VR_0 (\kappa ).
$$
We often use the formula 
\begin{equation}
R(\kappa )^* = -e^{i\overline{\kappa}} UR (\overline{\kappa}) ,
\label{S4_eq_resolventadjoint}
\end{equation}
which follows from $R(\kappa )^* = ((e^{i\overline{\kappa}} -U) e^{-i\overline{\kappa}} U^* )^{-1} $.

\begin{lemma}
For $ \theta \in [0,2\pi )$, we have $R(\theta \pm i0 )= \lim_{\epsilon \downarrow 0} R(\theta -i\log (1\mp \epsilon )) \in {\bf B} (\mathcal{B};\mathcal{B}^* )$ in the weak $*$ topology.
\label{S4_lem_LAP}
\end{lemma}

Proof.
Let $J$ be a compact interval in $[0,2\pi )$.
Due to the resolvent equation (\ref{S4_eq_resolventeq}) and Lemma \ref{S3_lem_LAP0}, there exists a constant $c>0$ such that 
\begin{equation}
\| R(\kappa )f\| _{\mathcal{B}^*} \leq c(\| f\| _{\mathcal{B}}+ \| VR(\kappa )f\| _{\mathcal{B}} ),
\label{S4_eq_LAP11}
\end{equation}
for $ \mathrm{Re} \, \kappa \in J $ and $\mathrm{Im} \, \kappa \not= 0 $.

First of all, let us prove that there exists a constant $c>0$ such that
\begin{equation}
\| R(\kappa )f\| _{\mathcal{B}^*} \leq c\| f\| _{\mathcal{B}} , \quad \mathrm{Re} \, \kappa \in J , \quad \mathrm{Im} \, \kappa \not= 0 .
\label{S4_eq_LAP12}
\end{equation}
Suppose that this inequality does not hold.
Without loss of generality, we can take sequences $\{ f^j \} _{j=1}^{\infty} \subset \mathcal{B} $ and $\{ \kappa_j \} _{j=1}^{\infty} \subset {\bf C} \setminus {\bf R} $ such that $\| R(\kappa_j )f^j \| _{\mathcal{B}^*} =1$, $\| f^j \| _{\mathcal{B}} \to 0$ and $\kappa_j \to \theta +i0 $ as $j\to \infty $. 
We put $ u^j = R(\kappa_j )f^j$.
Since the range of the operator $ V $ is a finite dimensional subspace of $\mathcal{B}$, there exists a subsequence $\{ u^{j_k} \} _{k=1}^{\infty}$ such that $ Vu^{j_k} $ converges in $\mathcal{B}$.
Now let $Vu^{j_k}$ converge to a sequence $g\in \mathcal{B}$.
It follows from the resolvent equation (\ref{S4_eq_resolventeq}) and Lemma \ref{S3_lem_LAP0} that 
$$
u:= \lim_{k\to \infty} u^{j_k} = \lim_{k\to \infty} \left( R_0 (\kappa_{j_k} )f^{j_k} -R_0 (\kappa_{j_k}) Vu^{j_k} \right) = -R_0 (\theta +i0 )g,
$$
in the weak $ *$ sense.
Then $u$ satisfies the equation $(U-e^{i\theta} )u=0 $ and $u=-R_0 ( \theta +i0 )g$ for $g\in \mathcal{B}$.
Applying Lemma \ref{S3_lem_uniqueness} and Corollary \ref{S3_cor_helmhltzunique}, we have $u=0$.
This is a contradiction in view of $\| u^{j_k} \| _{\mathcal{B}^* } =1 $ for all $k$.

Next we show the existence of the limit $R(\theta +i0 )$ in ${\bf B} (\mathcal{B}; \mathcal{B}^* )$ in the weak $*$ sense.
We consider a sequence $\{ \kappa_j \} _{j=1}^{\infty} $ where $\kappa_j = \theta -i\log (1\mp \epsilon_j ) $ with $\epsilon_j \downarrow 0 $.
For $f\in \mathcal{B}$, we put $u^j = R(\kappa_j )f$.
Due to the inequality (\ref{S4_eq_LAP12}), there exists a subsequence $ \{ u^{j_k} \} _{k=1}^{\infty} $ such that $Vu^{j_k}$ converges a sequence $g\in \mathcal{B}$ as above.
Then the limit $u:= \lim_{k\to \infty} u^{j_k} \in \mathcal{B}^*$ exists in the weak $*$ sense in view of 
$$
u= \lim_{k\to \infty} \left( R_0 (\kappa_{j_k} )f -R_0 (\kappa_{j_k}) Vu^{j_k} \right) =R_0 (\theta +i0)f -R_0 (\theta +i0 )g.
$$
The estimate
$$
\| g-Vu \| _{\mathcal{B}} \leq \| g-Vu^{j_k} \| _{\mathcal{B}} +\| V\| _{{\bf B} (\mathcal{B}^* ;\mathcal{B} )} \| u-u^{j_k}  \| _{\mathcal{B}^*} ,
$$
implies $ g=Vu$.
Let us prove that $u^j$ itself converges to $u$ in $\mathcal{B}^* $.
Assume that there exists another subsequence $\{ u^{j_l} \} _{l=1}^{\infty} $ such that $\lim _{l\to \infty} u^{j_l} =v\not= u$ in the weak $*$ sense for $v\in \mathcal{B}^* $.
We have 
$$
v=R_0 (\theta +i0 )f-R_0 (\theta +i0 )Vv,
$$
as above.
Then $u-v$ satisfies $(U-e^{i\theta} )(u-v) =0$ and $u-v= -R_0 (\theta +i0 )V(u-v)$.
Applying Lemma \ref{S3_lem_uniqueness} and Corollary \ref{S3_cor_helmhltzunique}, we obtain $u=v$.
This is a contradiction.

For $ R(\theta -i0 )$, the proof is similar.
\qed

\medskip

\textit{Remark.}
The limiting absorption principle in the sense of ${\bf B} (\mathcal{B};\mathcal{B}^* )$ was introduced by \cite{AgHo} (for self-adjoint partial differential operators with simple characteristics).
The pair $\mathcal{B}$-$\mathcal{B}^*$ is optimal for which the limiting absorption principle holds.
We also mention \cite{Ti} in which a rigorous proof of the limiting absorption principle for unitary operators was given as a general theory on Hilbert spaces.
Applying it for our case, we can see $R(\theta \pm i0)\in {\bf B} (\ell^{2,s},\ell^{2,-s})$ for any $s>1/2$ as a direct consequence.
However, since we adopt the framework of $\mathcal{B}$-$\mathcal{B}^* $ argument, we decided to rewrite a complete proof of our concrete argument for the sake of completeness of the paper.

\medskip

By the similar way, we can also prove the following lemma.

\begin{lemma}
Let $J$ be a compact interval in $ [0,2\pi )$.
The mapping $J\ni \theta \mapsto (R (\theta \pm i0 )f,g)$ for $f,g \in \mathcal{B}$ is continuous.
\label{S4_lem_continuity}
\end{lemma}

As a direct consequence of the resolvent equation (\ref{S4_eq_resolventeq}) and Lemmas \ref{S3_lem_LAP0} and \ref{S4_lem_LAP}, we obtain the uniqueness of the incoming (for $+$) or outgoing (for $-$) solution to the equation $(U-e^{i\theta} )u=f$ for $f \in \mathcal{B}$.

\begin{cor}
The solution $u^{(\pm )} \in \mathcal{B}^*$ to the equation $ (U -e^{i\theta} )u^{(\pm )} =f$ for $f\in \mathcal{B} $ is incoming (for $+$) or outgoing (for $-$) if and only if $u^{(\pm )}=R (\theta \pm i0 )f$.  
\label{S4_cor_helmhltzunique2}
\end{cor}


\subsection{Distorted Fourier transformation for QW}
Here we introduce the spectral representations for $U_0 $ and $U$.
The spectral representations are (distorted) Fourier transformations associated with $U_0$ and $U$.
Moreover, the generalized eigenfunctions are constructed by the spectral representations.

Let $ \mathcal{F}^{(0)} (\theta ) =( \mathcal{F}^{(0)}_{\leftarrow} (\theta ),  \mathcal{F}^{(0)} _{\rightarrow} (\theta ) ,  \mathcal{F}^{(0)} _{\downarrow} (\theta ) ,  \mathcal{F}^{(0)} _{\uparrow} (\theta )    )$ for $\theta \in [0,2\pi )$ be defined by
\begin{gather*}
( \mathcal{F}^{(0)} _{\leftarrow} (\theta )f)(x_2 ) = \frac{1}{\sqrt{2\pi}} \sum _{y_1 \in {\bf Z}} e^{-i\theta y_1} f_{\leftarrow} (y_1 ,x_2 ) {\bf e}_{\leftarrow} , \\
 ( \mathcal{F}^{(0)} _{\rightarrow} (\theta )f)(x_2 ) = \frac{1}{\sqrt{2\pi}} \sum _{y_1 \in {\bf Z}} e^{i\theta y_1} f_{\rightarrow} (y_1 ,x_2 ) {\bf e}_{\rightarrow} , \\
( \mathcal{F}^{(0)} _{\downarrow} (\theta )f)(x_1 ) = \frac{1}{\sqrt{2\pi}} \sum _{y_2 \in {\bf Z}} e^{-i\theta y_2} f_{\downarrow} (x_1 ,y_2 ) {\bf e}_{\downarrow} ,\\
( \mathcal{F}^{(0)} _{\uparrow} (\theta )f)(x_1 ) = \frac{1}{\sqrt{2\pi}} \sum _{y_2 \in {\bf Z}} e^{i\theta y_2} f_{\uparrow} (x_1 ,y_2 ) {\bf e}_{\uparrow} ,
\end{gather*}
for $f\in \mathcal{B}$.
Note that $\mathcal{F}^{(0)} (\theta ) \in {\bf B} (\mathcal{B};{\bf h} (\theta ))$ where the Hilbert space ${\bf h} (\theta )$ is defined by
$$
{\bf h} ( \theta )=\bigoplus _{p\in \{ \leftarrow ,\rightarrow, \downarrow ,\uparrow\} } \ell^2 ({\bf Z};{\bf C} ) {\bf e} _{p} ,
$$
with the inner product
$$
(\phi,\psi )_{{\bf h} (\theta )} = \sum_{ p\in \{ \leftarrow , \rightarrow \}} \sum_{x_2 \in {\bf Z}}  \phi_{p} (x_2 ) \overline{\psi_{p} (x_2 )} 
 + \sum_{ p\in \{ \downarrow , \uparrow \}} \sum_{x_1 \in {\bf Z}}  \phi_{p} (x_1 ) \overline{\psi_{p} (x_1 )} ,  
$$ 
for $\phi = \sum _{p\in \{ \leftarrow , \rightarrow , \downarrow , \rightarrow \}} \phi_p {\bf e} _p $ and $\psi = \sum_{p\in \{ \leftarrow , \rightarrow , \downarrow , \uparrow \}} \psi_p {\bf e}_p $.

\begin{lemma}
Let $f\in \mathcal{B}$.
We have 
\begin{gather}
\begin{split}
R_0 ( \theta +i0 )f  \simeq  & \, \sqrt{2\pi} F(x_1 ) e^{i\theta (x_1 -1 )} ( \mathcal{F}_{\leftarrow}^{(0)} (\theta ) f)(x_2 ) \\ 
& \, + \sqrt{2\pi} F(-x_1 ) e^{-i\theta (x_1 +1 )} ( \mathcal{F}_{\rightarrow}^{(0)} (\theta ) f)(x_2 ) \\
& \, + \sqrt{2\pi} F(x_2 ) e^{i\theta (x_2 -1 )} ( \mathcal{F}_{\downarrow}^{(0)} (\theta ) f)(x_1 ) \\ 
& \, + \sqrt{2\pi} F(-x_2 ) e^{-i\theta (x_2 +1 )} ( \mathcal{F}_{\uparrow}^{(0)} (\theta ) f)(x_1 ) ,
\end{split}
\label{S4_eq_asymptotic0+}
\end{gather}
and 
\begin{gather}
\begin{split}
R_0 ( \theta -i0 )f  \simeq  & \, -\sqrt{2\pi} F(-x_1 ) e^{i\theta (x_1 -1 )} ( \mathcal{F}_{\leftarrow}^{(0)} (\theta ) f)(x_2 ) \\ 
& \, - \sqrt{2\pi} F(x_1 ) e^{-i\theta (x_1 +1 )} ( \mathcal{F}_{\rightarrow}^{(0)} (\theta ) f)(x_2 ) \\
& \, - \sqrt{2\pi} F(-x_2 ) e^{i\theta (x_2 -1 )} ( \mathcal{F}_{\downarrow}^{(0)} (\theta ) f)(x_1 ) \\ 
& \, - \sqrt{2\pi} F(x_2 ) e^{-i\theta (x_2 +1 )} ( \mathcal{F}_{\uparrow}^{(0)} (\theta ) f)(x_1 ) .
\end{split}
\label{S4_eq_asymptotic0-}
\end{gather}
\label{S4_lem_asymptotic0}
\end{lemma}

In view of (\ref{S3_eq_resolvent0+}) and (\ref{S3_eq_resolvent0-}), we also have the following lemma.

\begin{lemma}
We have 
$$
( R_0 ( \theta +i0 )f - R_0 ( \theta -i0 )f,g) =2\pi e^{-i\theta} ( \mathcal{F}^{(0)} (\theta )f, \mathcal{F}^{(0)} (\theta )g )_{{\bf h} (\theta ) } ,
$$
for $ \theta \in [0,2\pi )$ and $f,g\in \mathcal{B}$.
\label{S4_lem_stone}
\end{lemma}

Now we have arrived at the formula of generalized eigenfunctions of $U_0$.
Taking the adjoint operator $ \mathcal{F}^{(0)} (\theta )^* \in {\bf B} ({\bf h} (\theta );\mathcal{B}^* )$, we have 
$$
(\mathcal{F}^{(0)} (\theta )^* \phi )(x)= \frac{1}{\sqrt{2\pi}} \left[ \begin{array}{c} e^{i\theta x_1} \phi _{\leftarrow} (x_2)  \\ e^{-i\theta x_1} \phi _{\rightarrow } (x_2 ) \\ e^{i\theta x_2} \phi _{\downarrow} (x_1 ) \\ e^{-i \theta x_2} \phi _{\uparrow} (x_1 ) \end{array} \right] , \quad \phi \in {\bf h} (\theta ).
$$

\begin{lemma}
For $ \theta \in [0,2\pi )$ and $\phi \in {\bf h} (\theta )$, $\mathcal{F}^{(0)} (\theta )^* \phi \in \mathcal{B}^* $ satisfies $(U_0 -e^{i\theta} ) \mathcal{F}^{(0)} (\theta )^* \phi =0 $.
\label{S4_lem_generalizedef0}
\end{lemma}

Let us turn to the distorted Fourier transformation associated with $U$.
Due to the resolvent equation (\ref{S4_eq_resolventeq}) and Lemma \ref{S3_lem_asymptotic0}, we define the operator $\mathcal{F}^{(\pm )} (\theta )=( \mathcal{F}^{(\pm )} _{\leftarrow} (\theta ), \mathcal{F}^{(\pm )} _{\rightarrow} (\theta ), \mathcal{F}^{(\pm )} _{\downarrow} (\theta ) , \mathcal{F}^{(\pm )} _{\uparrow} (\theta ) ) $ for $\theta \in [0,2\pi )$ by
$$
\mathcal{F}_p ^{(\pm )} (\theta )= \mathcal{F}_p^{(0)} (\theta ) (1-V R(\theta \pm i0 )), \quad p\in \{ \leftarrow , \rightarrow, \downarrow , \uparrow \} .
$$  

The operator $ \mathcal{F}^{(\pm)} (\theta )$ appears in the asymptotic behavior of $R(\theta \pm i0 )f$ at infinity.
The following lemma is a direct consequence of Lemma \ref{S4_lem_asymptotic0} and the resolvent equation (\ref{S4_eq_resolventeq}).

\begin{lemma}
Let $f\in \mathcal{B}$.
We have 
\begin{gather}
\begin{split}
R ( \theta +i0 )f  \simeq  & \, \sqrt{2\pi} F(x_1 ) e^{i\theta (x_1 -1 )} ( \mathcal{F}_{\leftarrow}^{(+)} (\theta ) f)(x_2 ) \\ 
& \, + \sqrt{2\pi} F(-x_1 ) e^{-i\theta (x_1 +1 )} ( \mathcal{F}_{\rightarrow}^{(+)} (\theta ) f)(x_2 ) \\
& \, + \sqrt{2\pi} F(x_2 ) e^{i\theta (x_2 -1 )} ( \mathcal{F}_{\downarrow}^{(+)} (\theta ) f)(x_1 ) \\ 
& \, + \sqrt{2\pi} F(-x_2 ) e^{-i\theta (x_2 +1 )} ( \mathcal{F}_{\uparrow}^{(+)} (\theta ) f)(x_1 ) ,
\end{split}
\label{S4_eq_asymptotic+}
\end{gather}
and 
\begin{gather}
\begin{split}
R ( \theta -i0 )f  \simeq  & \, -\sqrt{2\pi} F(-x_1 ) e^{i\theta (x_1 -1 )} ( \mathcal{F}_{\leftarrow}^{(-)} (\theta ) f)(x_2 ) \\ 
& \, - \sqrt{2\pi} F(x_1 ) e^{-i\theta (x_1 +1 )} ( \mathcal{F}_{\rightarrow}^{(-)} (\theta ) f)(x_2 ) \\
& \, - \sqrt{2\pi} F(-x_2 ) e^{i\theta (x_2 -1 )} ( \mathcal{F}_{\downarrow}^{(-)} (\theta ) f)(x_1 ) \\ 
& \, - \sqrt{2\pi} F(x_2 ) e^{-i\theta (x_2 +1 )} ( \mathcal{F}_{\uparrow}^{(-)} (\theta ) f)(x_1 ) .
\end{split}
\label{S4_eq_asymptotic-}
\end{gather}
\label{S4_lem_asymptotic}
\end{lemma}

An analogue of Lemma \ref{S4_lem_stone} holds.

\begin{lemma}
We have 
$$
( R ( \theta +i0 )f - R ( \theta -i0 )f,g) =2\pi e^{-i\theta} ( \mathcal{F}^{(\pm)} (\theta )f, \mathcal{F}^{(\pm)} (\theta )g )_{{\bf h} (\theta ) } ,
$$
for $ \theta \in [0,2\pi )$ and $f,g\in \mathcal{B}$.
\label{S4_lem_stone2}
\end{lemma}

Proof.
Note that the equalities 
\begin{gather}
\begin{split}
R( \theta & -i\log (1-\epsilon )) - R(\theta -i\log (1+\epsilon )) \\
&= -2\epsilon e^{i\theta}  R( \theta  -i\log (1-\epsilon )) R(\theta -i\log (1+\epsilon )) \\
&= -2 \epsilon e^{i\theta} R(\theta -i\log (1+\epsilon ))R( \theta  -i\log (1-\epsilon )) ,
\end{split}
\label{S4_eq_stone21}
\end{gather}
for $ \epsilon > 0$.
It follows from the second equality and the resolvent equation (\ref{S4_eq_resolventeq}) that
\begin{gather*}
\begin{split}
&R(\theta +i0 )-R(\theta -i0 ) \\
&= (1-R(\theta -i0)V) (R_0 (\theta +i0 )- R_0 (\theta -i0 ))(1-VR(\theta +i0 )).
\end{split}
\end{gather*}
From Lemma \ref{S4_lem_stone}, we have 
\begin{gather*}
\begin{split}
&(R(\theta +i0)f -R(\theta -i0)f,g) \\
&= 2\pi e^{-i\theta} (\mathcal{F}^{(0)} (\theta ) (1-VR(\theta +i0))f, \mathcal{F}^{(0)} (\theta )(1-V^* R(\theta -i0)^*)g) _{{\bf h} (\theta )} .
\end{split}
\end{gather*}
In view of (\ref{S4_eq_resolventadjoint}), we have $ R(\theta -i0)^* = -e^{i\theta} UR(\theta +i0)$.
Plugging this equality, $V^* U=-U_0^* V$ and $\mathcal{F}^{(0)} (\theta ) U_0^* = e^{-i\theta} \mathcal{F}^{(0)} (\theta )$, we obtain $\mathcal{F}^{(0)} (\theta )( 1-V^* R(\theta -i0 )^* )= \mathcal{F}^{(+)} (\theta )$.
We have proven the lemma for $ \mathcal{F}^{(+)} (\theta )$.
For $\mathcal{F}^{(-)} (\theta )$, we can prove the lemma by the same way, by using the first equality in (\ref{S4_eq_stone21}).
\qed

\medskip

By some direct computations, we can show that $\mathcal{F}^{(\pm)} (\theta )^* \in {\bf B} ({\bf h} (\theta ); \mathcal{B}^* )$ is an eigenoperator of $U$ as follows.

\begin{lemma}
For any $\phi \in {\bf h} (\theta )$, $\mathcal{F}^{(\pm)} (\theta )^* \phi \in \mathcal{B}^*$ satisfies the equation $ (U-e^{i\theta} ) \mathcal{F}^{(\pm)} (\theta )^* \phi =0 $.
$\mathcal{F}^{(\pm )} (\theta )^* \phi - \mathcal{F}^{(0 )} (\theta )^* \phi $ is outgoing (for $+$) or incoming (for $-$).
\label{S4_lem_eigenop}
\end{lemma}

Proof.
Note that 
\begin{gather}
\begin{split}
\mathcal{F}^{(\pm )} (\theta )^* &= \mathcal{F}^{(0)} (\theta )^* + e^{i\theta} UR(\theta \mp i0) V^* \mathcal{F} ^{(0)} (\theta )^* \\
&= \mathcal{F}^{(0)} (\theta )^* +e^{i\theta} \left( e^{i\theta} R( \theta \mp i0) V^* \mathcal{F}^{(0)} (\theta )^* + V^* \mathcal{F}^{(0)} (\theta )^*  \right) ,
\end{split}
\label{S4_eq_eigenop11}
\end{gather}
in view of (\ref{S4_eq_resolventadjoint}) and $(U-e^{i\theta} )R(\theta \pm i0 )g=g$ for any $g\in \mathcal{B}$.
Then we have $U\mathcal{F}^{(\pm)} (\theta )^* =e^{i\theta}  \mathcal{F}^{(\pm)} (\theta )^*$ by using the equalities $UV^* = -VU_0^*$ and $U_0^* \mathcal{F}^{(0)} (\theta )^* =e^{-i\theta}  \mathcal{F}^{(0)} (\theta )^*$.

The formula (\ref{S4_eq_eigenop11}) implies that $\mathcal{F}^{(\pm )} (\theta )^* \phi - \mathcal{F}^{(0 )} (\theta )^* \phi $ is outgoing (for $+$) or incoming (for $-$) in view of Lemma \ref{S4_lem_asymptotic} and the fact that $V^* \mathcal{F}^{(0)} (\theta )^* \phi =0$ at infinity.
\qed


\subsection{Characterization of generalized eigenfunction}
The set of solutions to the equation $(U-e^{i\theta} )u=0 $ for $u\in \mathcal{B}^* $ can be characterized by the range of $ \mathcal{F}^{(+)} (\theta )^* $.
To begin with, let us consider the case $ (U_0 -e^{i\theta} )u=0 $.

\begin{lemma}
We have $\mathrm{Ker} \mathcal{F}^{(0)} (\theta )^* = \{ 0 \} $.
The range of $ \mathcal{F}^{(0)} (\theta )^* $ is closed.
\label{S4_lem_closedrange0}
\end{lemma}

Proof.
The lemma follows from
$$
\| \mathcal{F}^{(0)} (\theta )^* \phi \|^2 _{\mathcal{B}^*} =\frac{1}{\pi} \| \phi \|^2 _{{\bf h} (\theta )} ,
$$
for any $ \phi \in {\bf h} (\theta )$.
\qed

\medskip

Due to Lemma \ref{S4_lem_closedrange0}, we can apply the following closed range theorem (see \cite[p. 205]{Yo}).

\begin{theorem}
Let $X_1$ and $X_2$ be Banach spaces and $(\cdot , \cdot )$ denote the pairing between $X_1$ or $X_2 $ and its dual spaces $X_1^*$ or $X_2^*$, respectively.
For $T\in {\bf B} (X_1 ;X_2 )$, the following assertions are equivalent.
\begin{enumerate}
\item $\mathrm{Ran} \, T $ is closed.

\item $\mathrm{Ran} \, T^* $ is closed.

\item $\mathrm{Ran} \, T  = (\mathrm{Ker} \, T^* )^{\perp} := \{ y\in X_2 \ ; \ (y,y^* )=0, \ \forall y^* \in \mathrm{Ker} \, T^* \} $.

\item $\mathrm{Ran} \, T ^*  = (\mathrm{Ker} \, T )^{\perp} := \{ x^* \in X_1^* \ ; \ (x,x^*)=0, \ \forall x \in \mathrm{Ker} \, T \} $.

\end{enumerate}
\label{S4_thm_closedrange}
\end{theorem}

Now we can prove the characterization of solutions to the equation $(U_0 -e^{i\theta} )u=0$ in $\mathcal{B}^*$ as follows.
We define the operator $\widehat{\mathcal{F}}^{(0)} (\theta )=(\widehat{\mathcal{F}}^{(0)} _{\leftarrow} (\theta ) , \widehat{\mathcal{F}}^{(0)}_{\rightarrow} (\theta ) , \widehat{\mathcal{F}}^{(0)}_{\downarrow} (\theta ) , \widehat{\mathcal{F}}^{(0)}_{\uparrow} (\theta ) )$ which is the Fourier transform of $\mathcal{F}^{(0)} (\theta )$ by
\begin{gather*}
(\widehat{\mathcal{F}}^{(0)} _{\leftarrow} (\theta ) f)( \xi _2 ) = \widehat{f} _{\leftarrow} (\theta , \xi_2 ) , \quad (\widehat{\mathcal{F}}^{(0)} _{\rightarrow} (\theta ) f)( \xi _2 ) = \widehat{f} _{\leftarrow} (-\theta , \xi_2 ) , \\
(\widehat{\mathcal{F}}^{(0)} _{\downarrow} (\theta ) f)( \xi _1 ) = \widehat{f} _{\downarrow} (\xi_1 , \theta ) , \quad (\widehat{\mathcal{F}}^{(0)} _{\uparrow} (\theta ) f)( \xi _1 ) = \widehat{f} _{\uparrow} (\xi_1 , -\theta ) , 
\end{gather*}
for $\xi = (\xi_1 , \xi_2 )\in {\bf T}^2 $.
Here we have identified ${\bf T}^2$ with $[-\pi ,\pi )^2 $.

\begin{lemma}
We have $\mathcal{F}^{(0)} (\theta ) \mathcal{B} = {\bf h} (\theta )$ and $\{ u\in \mathcal{B}^* \ ; \ (U_0 -e^{i\theta} )u=0 \} = \mathcal{F}^{(0)} (\theta )^* {\bf h} (\theta )$.
\label{S4_lem_chsolution0}
\end{lemma}

Proof.
We apply Theorem \ref{S4_thm_closedrange}, taking $X_1 = \mathcal{B}$, $X_2 = {\bf h } (\theta )$ and $T=\mathcal{F}^{(0)} (\theta )$.
The relation $\mathcal{F}^{(0)} (\theta ) \mathcal{B} = {\bf h} (\theta )$ follows from the assertion (3) in view of Lemma \ref{S4_lem_closedrange0}.
For the proof of the latter part, we have only to show $(u,f)=0$ when $u\in \mathcal{B}^*$, $f\in \mathcal{B}$, $(U-e^{i\theta} )u=0$ and $\mathcal{F}^{(0)} (\theta )f=0 $.
Since $\widehat{u} $ satisfies $(\widehat{U}_0 (\xi )-e^{i\theta} )\widehat{u} (\xi )=0$, we have $\mathrm{supp} \widehat{u}_{\leftarrow} \subset \{ \xi \in {\bf T}^2 \ ; \ x_1 = \theta \}$, $\mathrm{supp} \widehat{u}_{\rightarrow} \subset \{ \xi \in {\bf T}^2 \ ; \ x_1 = -\theta \}$, $\mathrm{supp} \widehat{u}_{\downarrow} \subset \{ \xi \in {\bf T}^2 \ ; \ x_2 = \theta \}$ and $\mathrm{supp} \widehat{u}_{\uparrow} \subset \{ \xi \in {\bf T}^2 \ ; \ x_2 =- \theta \}$.
It follows 
\begin{gather*}
\begin{split}
(u,f) &= \int _{{\bf T} } \left( \widehat{u} _{\leftarrow} ( \theta , \xi_2 ) \overline{\widehat{f} _{\leftarrow} (\theta , \xi_2 )} +    \widehat{u} _{\rightarrow} (- \theta , \xi_2 ) \overline{\widehat{f} _{\rightarrow} (-\theta , \xi_2 )} \right)  d\xi_2  \\
&\quad +  \int _{{\bf T} } \left( \widehat{u} _{\downarrow} ( \xi_1 , \theta ) \overline{\widehat{f} _{\downarrow} (\xi_1 , \theta  )} +    \widehat{u} _{\uparrow} (\xi_1 , - \theta  ) \overline{\widehat{f} _{\uparrow} (\xi_1 , -\theta )} \right)  d\xi_1 .  
\end{split}
\end{gather*}
In view of $\widehat{\mathcal{F}} ^{(0)} (\theta )f=0$, we obtain $ (u,f)=0$.
\qed

\medskip

Let us turn to the equation $(U-e^{i\theta} )u=0 $ in $ \mathcal{B}^* $.

\begin{theorem}
We have $\mathcal{F}^{(+)} (\theta ) \mathcal{B} = {\bf h} (\theta )$ and $\{ u\in \mathcal{B}^* \ ; \ (U -e^{i\theta} )u=0 \} = \mathcal{F}^{(+)} (\theta )^* {\bf h} (\theta )$.
\label{S4_thm_chsolution}
\end{theorem}

Proof.
By using the resolvent equation (\ref{S4_eq_resolventeq}), we have
\begin{gather}
\begin{split}
&(1-R(\theta +i0)^* V^* ) (1+R_0 (\theta +i0)^* V^* ) \\
&= (1+R_0 (\theta +i0)^* V^* )(1-R(\theta +i0)^* V^* )=1.
\end{split}
\label{S4_eq_chsolution11}
\end{gather}
Then $1-R (\theta +i0 )^* V^* $ is invertible.
It follows that $ \mathrm{Ker} \mathcal{F}^{(+)} (\theta )^* = \{ 0 \} $ and the range of $\mathcal{F}^{(+)} (\theta )^* $ is closed.
Applying the assertion (3) of Theorem \ref{S4_thm_closedrange}, we obtain $ \mathcal{F}^{(+)} (\theta ) \mathcal{B} = {\bf h} (\theta )$.

For the remaining part, we have only to prove $\{ u\in \mathcal{B}^* \ ; \ (U-e^{i\theta} )u=0 \} \subset \mathcal{F}^{(+)} (\theta )^* {\bf h} (\theta )$.
Suppose that $u\in \mathcal{B}^* $ satisfies $(U-e^{i\theta} )u=0 $.
We define $u^{(0)} \in \mathcal{B}^* $ by
\begin{gather}
\begin{split}
u^{(0)} &= (1+R_0 ( \theta +i0 )^* V^* )u \\
&= u-e^{i\theta} (1+e^{i\theta} R_0 ( \theta -i0 )) V^* u.
\end{split}
\label{S4_eq_chsolution12}
\end{gather}
In view of $U^* u= e^{-i\theta} u $ and $ U_0 V^* =- VU^* $, we have $ (U_0 -e^{i\theta} ) u^{(0)} =0 $. 
Lemma \ref{S4_lem_chsolution0} implies $u^{(0)} = \mathcal{F}^{(0)} (\theta )^* \phi $ for some $\phi \in {\bf h} (\theta )$.
Due to (\ref{S4_eq_chsolution12}), we have $(1+R_0 (\theta +i0 )^* V^* )u=\mathcal{F}^{(0)} (\theta )^* \phi $.
By using (\ref{S4_eq_chsolution11}), we obtain
$$
u= (1-R(\theta +i0 )^* V^* )\mathcal{F}^{(0)} (\theta )^* \phi = \mathcal{F}^{(+)} (\theta )^* \phi .
$$
Then $u\in \mathrm{Ran} \mathcal{F}^{(+)} (\theta )^* $.
\qed


\subsection{Spectral decomposition of scattering operator}
In view of Lemma \ref{S3_cor_absenceEV}, we have only to show the absence of $\sigma_{sc} (U)$ in order to prove the absolute continuity of $\sigma (U)$.
It is well-known that Stone's formula and the limiting absorption principle of the resolvent operator imply the absolute continuity of the essential spectrum of Schr\"{o}dinger operators.
An analogue of this argument holds for the $d$DQW as follows.

\begin{lemma}
We have $ \sigma_{sc} (U) = \emptyset $.
\label{S5_lem_singularconti}
\end{lemma}

Proof.
Let $ \theta \in (0,2\pi )$ and $ I= (\theta -\epsilon , \theta + \epsilon ) $ for small $ \epsilon > 0 $.
The lemma follows from Lemma \ref{S4_lem_continuity} and the formula (\ref{S3_eq_stone}) for $U$ :
\begin{equation}
(E_U ( ( \theta - \epsilon , \theta ' ) )f,f) _{\ell^2} = \int_{\theta - \epsilon}^{\theta '} \frac{e^{i\omega}}{2\pi} ( R ( \omega + i0 )f - R(\omega -i0 )f,f) d\omega , 
\label{S5_eq_stone}
\end{equation}
for $ \theta ' \in I $ and $ f\in \mathcal{B} $.
The proof is same as \cite[Lemma 4.6]{Mo}.
\qed

\medskip

Let the Hilbert space ${\bf H}$ be defined by ${\bf H} = L^2 ([0,2\pi ); {\bf h} (\theta ); d\theta )$ with the inner product
$$
(f,g) _{{\bf H}} = \int _0^{2\pi} (f(\theta ), g(\theta )) _{{\bf h} (\theta )} d\theta .
$$
The operators $ \mathcal{F}^{(0)} $ and $ \mathcal{F}^{(\pm)} $ are defined by 
$$
( \mathcal{F}^{(0)} f)( \theta )= \mathcal{F}^{(0)} (\theta )f, \quad (\mathcal{F}^{(\pm)} f)( \theta )= \mathcal{F}^{(\pm)} (\theta )f,
$$
for $\theta \in [0,2\pi )$ and $f\in \mathcal{B}$.
The following lemma follows from the definition of $\mathcal{F}^{(0)} (\theta )$.

\begin{lemma}
The following assertions holds.
\begin{enumerate}
\item The operator $\mathcal{F}^{(0)} $ can be extended uniquely to a unitary operator from $\ell^2 $ to ${\bf H}$.

\item We have $ ( \mathcal{F}^{(0)} U_0 f)(\theta )= e^{i\theta} (\mathcal{F}^{(0)} f)(\theta )$ and $ ( \mathcal{F}^{(0)} U_0^* f)(\theta )= e^{-i\theta} (\mathcal{F}^{(0)} f)(\theta )$ for $ \theta \in [0,2\pi )$ and $f\in \ell^2 $.
\end{enumerate}
\label{S5_lem_F0}
\end{lemma}

The operator $ \mathcal{F}^{(\pm )}$ also diagonalizes $U$ as follows.

\begin{lemma}
The following assertions holds.
\begin{enumerate}
\item The operator $\mathcal{F}^{(\pm)} $ can be extended uniquely to a unitary operator from $\ell^2 $ to ${\bf H}$.

\item We have $ ( \mathcal{F}^{(\pm)} U f)(\theta )= e^{i\theta} (\mathcal{F}^{(\pm)} f)(\theta )$ and $ ( \mathcal{F}^{(\pm)} U^* f)(\theta )= e^{-i\theta} (\mathcal{F}^{(\pm)} f)(\theta )$ for $ \theta \in [0,2\pi )$ and $f\in \ell^2 $.
\end{enumerate}
\label{S5_lem_Fpm}
\end{lemma}

Proof.
Lemma \ref{S4_lem_stone2} and the formula (\ref{S5_eq_stone}) imply
$$
(E_U ((\theta _1 , \theta _2 ))f,g)_{\ell^2} = \int _{\theta_1} ^{\theta_2} ( \mathcal{F}^{(\pm )} (\theta )f , \mathcal{F}^{(\pm)} (\theta )g) _{{\bf h} (\theta )} d\theta , 
$$
for $\theta_1 < \theta_2 $, $\theta_1 , \theta_2 \in [0,2\pi )$.
It follows that $ \mathcal{F}^{(\pm )} $ is a partial isometry from $\mathcal{H}_{ac} (U)$ to ${\bf H}$.
In view of Corollary \ref{S3_cor_absenceEV} and Lemma \ref{S5_lem_singularconti}, we have $\mathcal{H}_p (U)= \mathcal{H}_{sc} (U)= \emptyset $.
Then we obtain the assertion (1).

The assertion (2) follows from the definition of $ \mathcal{F}^{(\pm)} $.
The details of computation is same as \cite[Theorem 4.7]{Mo}.
\qed

\medskip

The Fourier transform of the scattering operator $\Sigma = W_+^* W_- $ is defined by 
\begin{equation}
\widehat{\Sigma} = \mathcal{F}^{(0)} \Sigma (\mathcal{F}^{(0)})^* .
\label{S4_eq_spectraltransS}
\end{equation}
By the same way of \cite[Theorem 5.3]{Mo}, we can see that $\widehat{\Sigma}$ can be decomposed as 
$$
\widehat{\Sigma} = \int _0^{2\pi} \oplus \widehat{\Sigma} (\theta )d\theta ,
$$
and $ \widehat{\Sigma} (\theta )$ satisfies the following properties.

\begin{theorem}
Let $ \theta \in [0,2\pi )$.
The S-matrix $\widehat{\Sigma} ( \theta )$ satisfies the following assertions.
\begin{enumerate}
\item For $f\in {\bf H}$, we have $ ( \widehat{\Sigma} f )( \theta )= \widehat{\Sigma} (\theta )f(\theta ) $.

\item $ \widehat{\Sigma} (\theta )$ is unitary on ${\bf h} (\theta )$.

\item We have $ \widehat{\Sigma} (\theta )= 1-2\pi e^{i\theta} A(\theta )$ where
$$
A(\theta )=( A_{\leftarrow} (\theta ), A_{\rightarrow} (\theta ), A_{\downarrow} (\theta ), A_{\uparrow} (\theta )) ,
$$
is defined by 
$$
A_p (\theta )= \mathcal{F}^{(-)} _p (\theta ) V^* \mathcal{F}^{(0)} (\theta )^* ,
$$
for $p\in \{ \leftarrow , \rightarrow , \downarrow , \uparrow \} $.
\end{enumerate}
\label{S5_thm_Smatrix}
\end{theorem}

\textit{Remark.}
Usually the statement of the S-matrix holds for $\mathrm{a.e.} \, \theta $ in the continuous spectrum.
For the case where $\sigma _{ac} (U_0)$ has band gaps, the endpoints of $\sigma_{ac} (U_0) $ are the exceptional points.
If $U_0$ or $U$ have some eigenvalues embedded in the continuous spectrum, these embedded eigenvalues are also exceptional points. 
However, the spectrum of the free QW $U_0$ does not band gaps and its structure is uniform on the unit circle.
Indeed, there is no threshold in $ \theta$ for the limiting absorption principle (formulas (\ref{S3_eq_resolvent0+}) and (\ref{S3_eq_resolvent0-}), Lemmas \ref{S3_lem_LAP0} and \ref{S4_lem_LAP}, formulas (\ref{S4_eq_asymptotic+}) and (\ref{S4_eq_asymptotic-})) and the definition of the Hilbert space ${\bf h} (\theta )$.
Mathematically, the homogeneity of $\sigma_{ac} (U_0)$ follows from $ \frac{\partial }{\partial \xi_j} (e^{\pm i \xi_j} -e^{i\theta} )\big| _{\pm \xi_j =\theta} \not= 0$ for $j=1,2 $.
Due to the assumption for $C(x)$ and Corollary \ref{S3_cor_absenceEV}, there is no eigenvalue in the continuous spectrum.
Thus the fiber operator $\widehat{\Sigma} (\theta )$ can be defined for all $\theta$ modulo $2\pi$ without thresholds.
If we remove the second assumption for $C(x) $, the operator $U$ may have some eigenvalues with eigenfunctions with finite supports.
In this case, Theorem \ref{S5_thm_Smatrix} holds for $ \theta $ except for a finite number of embedded eigenvalues.

\medskip

Due to the representation (\ref{S4_eq_eigenop11}) of generalized eigenfunctions of $U$, we have for $u^{(+)} = \mathcal{F}^{(+)} (\theta )^* \phi $ and $u^{(0)} = \mathcal{F}^{(0)} (\theta )^* \phi $, $\phi \in {\bf h} (\theta )$,
\begin{gather}
\begin{split}
u^{(+)} - u^{(0)} \simeq & \, -\sqrt{2\pi} e^{i\theta} F(-x_1 ) e^{i\theta x_1} (A_{\leftarrow} (\theta ) \phi )(x_2) \\
& \, -\sqrt{2\pi} e^{i\theta} F(x_1 ) e^{-i\theta x_1} (A_{\rightarrow} (\theta )\phi )(x_2 ) \\
& \, -\sqrt{2\pi} e^{i\theta} F(-x_2 ) e^{i\theta x_2} (A_{\downarrow} (\theta ) \phi )(x_1) \\
& \, -\sqrt{2\pi} e^{i\theta} F(x_2 ) e^{-i\theta x_2} (A_{\uparrow} (\theta )\phi )(x_1 ). 
\end{split}
\label{S4_eq_asymptoticef+}
\end{gather}
In particular, the generalized eigenfunction $\mathcal{F}^{(+)} (\theta )^* \phi $ has the asymptotics
\begin{gather*}
(\mathcal{F}^{(+)} (\theta )^* \phi )_{\leftarrow} (x) = \frac{e^{i\theta x_1}}{\sqrt{2\pi}} (\widehat{\Sigma} _{\leftarrow} (\theta ) \phi )(x_2 )+o(1), \quad x_1 \to -\infty , \\
(\mathcal{F}^{(+)} (\theta )^* \phi )_{\rightarrow} (x) = \frac{e^{-i\theta x_1}}{\sqrt{2\pi}} (\widehat{\Sigma} _{\rightarrow} (\theta ) \phi )(x_2 )+o(1), \quad x_1 \to \infty ,
\end{gather*}
for every fixed $x_2 \in {\bf Z}$, and
\begin{gather*}
(\mathcal{F}^{(+)} (\theta )^* \phi )_{\downarrow} (x) = \frac{e^{i\theta x_2}}{\sqrt{2\pi}} (\widehat{\Sigma} _{\downarrow} (\theta ) \phi )(x_1 )+o(1), \quad x_2 \to -\infty , \\
(\mathcal{F}^{(+)} (\theta )^* \phi )_{\uparrow} (x) = \frac{e^{-i\theta x_2}}{\sqrt{2\pi}} (\widehat{\Sigma} _{\uparrow} (\theta ) \phi )(x_1 )+o(1), \quad x_2 \to \infty ,
\end{gather*}
for every fixed $x_1 \in {\bf Z}$.
Moreover, $v^{(+)} := u^{(+)} - u^{(0)}$ satisfies not only the asymptotics (\ref{S4_eq_asymptoticef+}) but equalities as follows :
\begin{gather*}
v^{(+)}_{\leftarrow} (x) = -\sqrt{2\pi} e^{i\theta} e^{i\theta x_1} (A_{\leftarrow} (\theta ) \phi )(x_2), \quad x_1 \leq -n_0 -1 , \\
v^{(+)}_{\rightarrow} (x) = -\sqrt{2\pi} e^{i\theta} e^{-i\theta x_1} (A_{\rightarrow} (\theta ) \phi )(x_2), \quad x_1 \geq n_0 +1 , \\
v^{(+)}_{\downarrow} (x) = -\sqrt{2\pi} e^{i\theta} e^{i\theta x_2} (A_{\downarrow} (\theta ) \phi )(x_1), \quad x_2 \leq -n_0 -1 , \\
v^{(+)}_{\uparrow} (x) = -\sqrt{2\pi} e^{i\theta} e^{-i\theta x_2} (A_{\uparrow} (\theta ) \phi )(x_1), \quad x_2 \geq n_0 +1 .
\end{gather*}
In fact, $v^{(+)}$ satisfies $ (U_0 -e^{i\theta} )v^{(+)} =0$ in ${\bf Z}^2 \setminus (D\cup \partial D)$.
Then we can see these equalities by the same argument of the proof of Lemma \ref{S3_lem_uniqueness}, considering the asymptotics (\ref{S4_eq_asymptoticef+}).

\begin{figure}[t]
\centering
\includegraphics[bb=0 0 384 414, width=4cm]{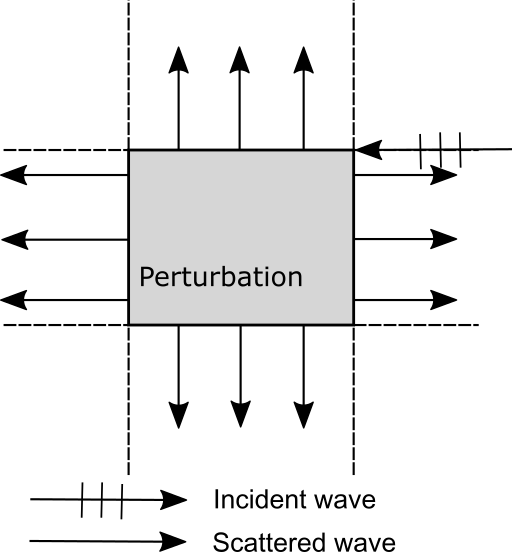}
\caption{The scattered wave with an incident wave which consists of one plane wave of the chirality $\leftarrow$.} 
\label{fig_ti_scattering2}
\end{figure}

For 2DQW, the scattered wave does not spread radially as above.
Since $V=U-U_0$ is finite rank, the scattered wave for every chirality passes along corridors. See Figure \ref{fig_ti_scattering2}.

\begin{theorem}
Let $\phi \in {\bf h} (\theta )$.
 \begin{enumerate}
\item $( A _{\leftarrow} (\theta ) \phi )(x_2 )$ and $ ( A _{\rightarrow} (\theta ) \phi )(x_2 )$ vanish for $|x_2 |\geq n_0 +1 $.

\item $( A_{\downarrow} (\theta ) \phi )(x_1 )$ and $ ( A _{\uparrow} (\theta ) \phi )(x_1 )$ vanish for $|x_1 |\geq n_0 +1 $.

\end{enumerate}
As a consequence, $ A (\theta )$ is an operator of finite rank.
\label{S4_thm_corridor}
\end{theorem}

Proof.
Note that 
$$
v^{(+)} = u^{(+)} - u^{(0)} 
= e^{i\theta} \left( e^{i\theta} R_0 (\theta -i0 )g +V^* u^{(0)} \right) ,
$$
where $g=V^* u^{(0)} - V R(\theta -i0) V^* u^{(0)} $ for $u^{(+)} = \mathcal{F}^{(+)} (\theta )^* \phi $ and $u^{(0)} = \mathcal{F}^{(0)} (\theta )^* \phi $, $\phi \in {\bf h} (\theta )$. 
In view of the assumption for $C(x)$, we have $(V^* u^{(0)} ) _{\leftarrow} (x) = ((C^* -1) S^{-1} u^{(0)})_{\leftarrow} (x)=0$ for $|x_2 | \geq n_0$. 
Similarly, we have $ (V R(\theta -i0) V^* u^{(0)})_{\leftarrow} (x ) = (S(C-1)R(\theta -i0) V^* u^{(0)})_{\leftarrow} (x)=0 $ for $|x_2 | \geq n_0$. 
Thus the formula (\ref{S3_eq_resolvent0-}) shows that $ v^{(+)} (x)$ vanishes for $|x_2 | \geq  0$.
This implies that $ ( A_{\leftarrow} (\theta )\phi ) (x_2 )=0 $ for $|x_2 | \geq 0 $.
The proofs for $A_p (\theta )\phi$ with $p\in \{ \rightarrow , \downarrow , \uparrow \} $ are similar.
\qed

\end{document}